\documentclass[usenatbib,useAMS]{mn2e}

\usepackage{graphicx,times}

\newcommand{\Msun}{\ensuremath{M_{\odot}}}
\newcommand{\Ha}{\ensuremath{{\rm H}_{\alpha}}}
\newcommand{\Hb}{\ensuremath{{\rm H}_{\beta}}}

\title[Spectroscopic properties of SDSS-III/BOSS galaxies]{Stellar velocity dispersions and emission line properties of SDSS-III/BOSS galaxies}

\author[Thomas et al.] {
\parbox[h]{\textwidth}{D.~Thomas$^{1,2}$, O.~Steele$^1$, C.~Maraston$^{1,2}$, J.~Johansson$^{1,3}$, A.~Beifiori$^{1,4}$, J.~Pforr$^{1,5}$, G.~Str\"omb\"ack$^1$, C.~A.~Tremonti$^6$, D.~Wake$^7$, 
D.~Bizyaev$^8$,
A.~Bolton$^9$,
H.~Brewington$^8$,
J.~R.~Brownstein$^{9}$,
J.~Comparat$^{10}$,
J.-P. Kneib$^{10}$,
E.~Malanushenko$^8$,
V.~Malanushenko$^8$,
D.~Oravetz$^8$,
K.~Pan$^8$,
J.~K.~Parejko$^{7}$,
D.~P.~Schneider$^{11,12}$,
A.~Shelden$^8$,
A.~Simmons$^8$,
S.~Snedden$^8$,
M.~Tanaka$^{13}$,
B.~A.~Weaver$^{14}$,
R.~Yan$^{14}$}
\vspace*{8pt}\\ 
$^1$Institute of Cosmology and Gravitation, University of Portsmouth, Dennis Sciama Building, Burnaby Road, Portsmouth, PO1 3FX, UK\\
$^2$SEPnet, South East Physics Network, (www.sepnet.ac.uk)\\
$^3$Max-Planck-Institut f\"ur Astrophysik, D-85748 Garching, Germany\\
$^{4}$Max-Planck-Institut f\"ur extraterrestrische Physik, Giessenbachstra§e, D-85748 Garching, Germany\\
$^{5}$NOAO, 950 North Cherry Ave., Tucson, AZ 85719, USA\\
$^6$Department of Astronomy, University of Wisconsin-Madison, 1150 University Ave, Madison, WI 53706, USA\\
$^7$Yale Center for Astronomy and Astrophysics, Yale University, New Haven, CT, USA\\
$^{8}$Apache Point Observatory, P.O. Box 59, Sunspot, NM 88349-0059, USA\\
$^9$Department of Physics and Astronomy, University of Utah, 115 South 1400 East, Salt Lake City, UT 84112, USA\\
$^{10}$Laboratoire dÕAstrophysique de Marseille, CNRS/Aix-Marseille Universit\'{e}, 38 rue Fr\'{e}d\'{e}ric Joliot-Curie, 13388, Marseille cedex 13, France\\
$^{11}$Department of Astronomy and Astrophysics, The Pennsylvania State University,
  University Park, PA 16802\\
$^{12}$Institute for Gravitation and the Cosmos, The Pennsylvania State University,
  University Park, PA 16802\\
$^{13}$Institute for the Physics and Mathematics of the Universe, The University of Tokyo, 5-1-5 Kashiwanoha, Kashiwa-shi, Chiba 277-8583, Japan\\
$^{14}$Center for Cosmology and Particle Physics, Department of Physics, New York University, New York, NY, 10003}

\date{Accepted ... Received 18 January 2013 ; in original form 23 July 2012}

\pagerange{\pageref{firstpage}--\pageref{lastpage}}
\pubyear{2012}

\begin{document}

\bibliographystyle{mn2e}
\maketitle

\label{firstpage}

\begin{abstract}
We perform a spectroscopic analysis of 492,450 galaxy spectra from the first two years of observations of the Sloan Digital Sky Survey-III/Baryonic Oscillation Spectroscopic Survey (BOSS) collaboration. This data set has been released in the ninth SDSS data release, the first public data release of BOSS spectra. We show that the typical signal-to-noise ratio of BOSS spectra, despite being low, is sufficient to measure stellar velocity dispersion and emission line fluxes for individual objects. We show that the typical velocity dispersion of a BOSS galaxy is $\sim 240$ km s$^{-1}$. The typical error in the velocity dispersion measurement is 14 per cent, and 93 per cent of BOSS galaxies have velocity dispersions with an accuracy of better than 30 per cent. The distribution in velocity dispersion is redshift independent between redshifts 0.15 and 0.7, which reflects the survey design targeting massive galaxies with an approximately uniform mass distribution in this redshift interval. We show that emission lines can be measured on BOSS spectra. However, the majority of BOSS galaxies lack detectable emission lines, as is to be expected because of the target selection design toward massive galaxies. {\bf\rm\rm\rm We analyse the emission line properties and present diagnostic diagrams using the emission lines [OII], H$\beta$, [OIII], H$\alpha$, and [NII] (detected in about 4 per cent of the galaxies) to separate star-forming objects and AGN. We show that the emission line properties are strongly redshift dependent and that there is a clear correlation between observed frame colours and emission line properties. Within in the low-$z$ sample (LOWZ) around $0.15<z<0.3$, half of the emission-line galaxies have LINER-like emission line ratios, followed by Seyfert-AGN dominated spectra, and only a small fraction of a few per cent are purely star forming galaxies. AGN and LINER-like objects, instead, are less prevalent in the high-$z$ sample (CMASS) around $0.4<z<0.7$, where more than half of the emission line objects are star forming. This is a pure selection effect caused by the non-detection of weak \Hb\ emission lines in the BOSS spectra.} Finally, we show that star forming, AGN and emission line free galaxies are well separated in the $g-r$ vs $r-i$ target selection diagram.
\end{abstract}



\section{Introduction}
\label{intro}
Within the framework of hierarchical galaxy formation theory \citep{WR78,Freetal85}, the formation and evolution of galaxies is driven by the interplay between dark matter and baryon physics involving complex processes such as gas accretion, star formation, black hole accretion, chemical enrichment, galactic winds etc. Progress in our understanding of galaxy formation crucially relies on the observational constraints that can be set on theoretical models. Traditionally, this can be done either through detailed studies of local galaxies \citep[e.g.][]{Kuntschner00,Davies01,Thomas05,KK04,Kuntschner10,Kormendy09,Kormendy10}, or through the analysis of remote galaxies at large distances using the universe as a time machine \citep[e.g.][]{Bundy06,Cimatti04,Genzel03a}.

Recent large galaxy surveys have opened a new avenue to approach this problem, by providing large data sets that allow statistical studies of large galaxy samples in the nearby universe. In the last decade an overwhelming number of studies based on data from the Sloan Digital Sky Survey \citep[][SDSS]{York00} have led to significant progress in our understanding of the local galaxy population \citep[e.g.][]{Blanton03,Kauffmann03a,Kauffmann03b,Tremonti04,Brinchmann04,Wake06,Panter07,Schawinski07b,Thomas10}. The Baryon Oscillation Spectroscopic Survey (hereafter BOSS), part of the Sloan Digital Sky Survey III collaboration \citep[][SDSS-III]{Eisenstein11}, will now extend this database to higher redshifts obtaining spectra of 1.5 million luminous galaxies up to redshifts $z\sim 0.7$ by 2014. This will allow large-scale statistical studies of the galaxy population at an epoch when the universe had only half of its current age.

In this paper we present the spectroscopic analysis of 492,450 galaxy spectra observed in the first 1.5 years of BOSS operation covering about one third of the total survey area. We introduce the spectroscopic analysis tools developed for this purpose and present stellar velocity dispersions, emission line fractions, and emission line classifications. The aim of this work is to characterise BOSS galaxies and to discuss their basic spectroscopic properties in the BOSS targeting colour-colour and colour-magnitude space.

The paper is organised as follows. We briefly introduce the BOSS project in Section~\ref{sec:data}. Section~\ref{sec:tool} presents the methodology and the calibration of the analysis tool. Results are shown in Sections~\ref{sec:velocity}, \ref{sec:emission} and \ref{sec:target}. The paper concludes with Section~\ref{sec:conclusions}.

\section{Data}
\label{sec:data}
BOSS is one of four surveys of the SDSS-III collaboration using an upgrade of the multi-object spectrograph on the 2.5m SDSS telescope \citep{Gunn06} to collect spectra of galaxies and quasars over 10,000 deg$^{2}$ on the sky. BOSS has started operation in Fall 2009 and by 2014 it will have observed about 1.5 million luminous galaxies up to redshifts $z\sim 0.7$. A comprehensive overview of the project is presented in \citet{Eisenstein11} and \citet{Dawson12}. {\bf\rm\rm The first public release of BOSS spectra in July 2012 as part of the ninth SDSS data release (DR9) is described in \citet{SDSSDR9}.}

\subsection{Target selection}
Galaxies and quasars are dealt with separately in the target selection for BOSS spectroscopy. Quasar selection is described in \citet{Ross12}, while the galaxy target selection algorithm is explained in detail in \citet{Dawson12}, and we refer the reader to these papers for a comprehensive description. Here we summarise the major aspects of galaxy target selection that are relevant for the present paper. Targets are selected from SDSS imaging \citep{Fukugita96,Gunn98} from the eighth data release \citep[DR8][]{SDSSDR8}. Galaxy target selection is based on the two colours $g-r$ and $r-i$, which have been shown to provide a powerful colour-colour space to select luminous galaxies up to redshift $z\sim 0.7$ in the construction of the SDSS luminous red galaxies (LRG) sample \citep{Eisenstein01} and in the 2dF-SDSS LRG and QSO survey \citep[][2SLAQ]{Cannon06}. BOSS reaches significantly deeper than the SDSS LRG sample and is significantly wider than 2SLAQ.

The galaxy sample is split in a high redshift sample called CMASS and a low redshift sample LOWZ. 
The limiting apparent magnitude for galaxy targets is $17.5<i<19.9\;$mag AB \citep{1983ApJ...266..713O} in the CMASS and $16<r<19.6\;$mag in the LOWZ samples. While the SDSS MAIN sample is purely magnitude limited, CMASS and LOWZ employ colour cuts in order to target higher-$z$ samples without being overwhelmed by low-$z$ contamination. These cuts are designed to target massive passively evolving galaxies, but also include some star-forming galaxies as explored in the present paper. The $r-i$ colour serves as a powerful and simple redshift estimator. At redshift $z\sim 0.4$ the $4000\;$\AA\ break moves from the $g$ into the $r$-band, which results in a steep rise of $r-i$ as a function of redshift beyond $z=0.4$. The two cuts, LOWZ and CMASS, reflect this dividing line at $z=0.4$. The model trajectory of a passively evolving massive galaxy with $M\sim 10^{11}\;$\Msun\ \citep{Maraston09b} shows that this galaxy class hits the detection limit of $i=19.9\;$mag at a redshift of $z\sim 0.7$. As a consequence, the CMASS sample contains massive galaxies in the redshift interval $0.4\leq z\leq 0.7$. The number density of BOSS galaxies drops sharply beyond $z=0.7$ \citep{Padmanabhan10,White11}. About one third of the LOWZ sample is covered with existing SDSS-I/II spectroscopy, while most of the CMASS galaxy spectra are obtained for the first time with BOSS observations. {\bf\rm\rm Note that we do not include LOWZ or CMASS targets with existing spectra from SDSS-I/II in the present analysis.}

A subsample called SPARSE has been defined within CMASS in order to allow us to study galaxies beyond the nominal CMASS selection cuts. The SPARSE sample contains galaxies with slightly bluer $r-i$ colours for targets with an $i$-band magnitude fainter than 19.5 \citep[for details see][]{Dawson12}. Furthermore, the BOSS dataset includes a sample obtained during commissioning called COMM that is part of the CMASS sample.  The current study includes the SPARSE and the COMM samples as part of CMASS.

Galaxy target selection is designed to construct a sample of massive galaxies with constant co-moving number density between redshifts 0.1 and 0.7 with a uniform mass distribution at all redshifts. The analyses of number density \citep{Padmanabhan10,White11} and galaxy masses \citep{Chen12,Maraston12} from early BOSS data show that this goal has been achieved. Both these constraints are critical for the proper measurement of large-scale structure as a function of redshift and the cosmology science goals of BOSS \citep{Anderson12}. The uniform mass distribution is crucial for galaxy evolution studies, as it provides a sample that is homogeneous in mass, thus minimising the mass bias when studying galaxy properties as a function of redshift.

\subsection{Spectroscopy}
BOSS multi-object spectroscopy is being performed with an upgrade of the SDSS  fibre-fed spectrograph. The hardware upgrade is introduced in \citet{Bhardwaj10}, \citet{Roe10}, and \citet{Smee12}, while details about the spectroscopic observations, data reduction, and the spectroscopic pipeline are provided in \citet{Bolton12}. The most important features of the upgrade include the increase of the number of  fibres to 1,000 per plate,  the decrease of  fibre diameter to 2'', and the employment of a new CCD with extended wavelength coverage to span 3600 to $10,000\;$\AA\ and much improved throughput. Large format VPH gratings and thick, fully-depleted CCDs are special features of the upgrade \citep{Roe10}. The SDSS pipeline is used for data reduction and simple measurements. Standard stars are targeted on each plate, so that the BOSS spectra have been flux-calibrated.

The upgrade implies that spectra of reasonable signal-to-noise ratio can be obtained in 1-hour exposures down to the limiting magnitude in the $i$-band of 19.9~mag. The typical signal-to-noise ratios of a BOSS galaxy spectrum are about 5 per \AA, which is sufficient for an accurate redshift determination. In fact the desired redshift efficiency of 95 per cent has been accomplished \citep{Ross11b,White11,Masters11}. The quality of the spectra is sufficient for the detection of emission lines and the measurement of simple dynamical quantities such as stellar velocity dispersion, which are subject of the present paper. The spectral resolution is $R\sim 2,000$, hence high enough for an accurate derivation of these basic quantities.

The typical signal-to-noise ratio clearly does not allow for more detailed spectroscopic studies of individual objects. However, very high signal-to-noise ratio spectra can be obtained through stacking. This approach is necessary for absorption line measurements and studies of element abundance ratios and other stellar population parameters, which is subject of future work. 

\subsection{The present sample}
BOSS began in December 2009, and has taken 534,474 galaxy spectra in total by July 2011, observing 813 plates over $3,275\;$deg$^{2}$ on the sky, which corresponds to about one third of the final survey footprint. We use this data set in the present work, which corresponds to the data volume that has been released in DR9, the first public data release of BOSS spectra. We select those objects that are classified as galaxies and for which the ZWARNING\_NQSO flag is set zero (reliable redshifts and excluding QSOs) resulting in 492,450 unique objects (excluding repeat observations). This includes all BOSS galaxies from DR9 that have been selected through the LOWZ and CMASS selection cuts. The SPARSE sample is included within the CMASS selection. We do not consider LOWZ or CMASS targets with spectra from SDSS I/II.

\section{Analysis tools}
\label{sec:tool}
Following the approaches of \citet{Schawinski07b} and \citet{Thomas10}, we use the publicly available codes pPXF \citep{CE04} and GANDALF v1.5 \citep{Sarzi06} to calculate stellar kinematics and to derive emission line properties. We refer to those papers for more information. In brief, GANDALF fits stellar population and Gaussian emission line templates to the galaxy spectrum {\em simultaneously} to separate stellar continuum and absorption lines from the ionized gas emission. Stellar kinematics are evaluated by pPXF. The subtraction of the emission line spectrum from the observed one produces a clean absorption line spectrum free from emission line contamination. This spectrum will be used for any further analysis of absorption features. The fits account for the impact of diffuse dust in the galaxy on the spectral shape adopting a \citet{Calzetti01} obscuration curve. The code further determines the kinematics of the gas (velocity and velocity dispersion) and measures emission line fluxes and equivalent widths (EWs) on the resulting Gaussian emission line template.

{\bf\rm\rm To calibrate our analysis tool we present a comparison with low-redshifts data from the previous surveys SDSS-I/II. We have selected a subset of $\sim 10,000$ SDSS galaxies from 20 randomly selected plates. Quantities published in Data Release 7 \citep[DR7][]{SDSSDR7} are adopted for the comparison. This is for calibration purposes only, and the DR7 set of galaxies is not part of the higher-redshift BOSS sample from DR9 analysed in this paper.}

\subsection{Stellar population templates}
The stellar population template is a critical ingredient in this procedure. We adopt the new stellar population models from \citet{Mastro11} based on the MILES stellar library \citep{Sanchez06a}. The wavelength range covered by these templates is $3,500-7,500\;$\AA. As this blue limit corresponds to a relatively red rest-frame wavelength of $\sim 5,200\;$\AA\ around $z\sim 0.5$, \citet{Mastro11} have augmented their templates with theoretical spectra at wavelengths $\lambda<3,500\;$\AA\ using the model by \citet{Maraston09a} based on the theoretical library UVBLUE \citep{Rodriguez05}. We use this hybrid model in the present study.

A proper stellar population fit requires a large set of templates spanning a wide range in the parameter space of formation ages, star formation histories, metallicities, stellar initial mass functions and dust attenuations \citep{Maraston06,Tojeiro09,Maraston10}. Such an analysis is computationally expensive. However, for the purpose of this work, we merely need the stellar population fits to separate the absorption spectrum from the emission spectrum, and to perform a simple analysis of stellar kinematics. To this end it is most critical to obtain a good representation of the absorption spectrum.  Hence in this case we can benefit from the age-metallicity degeneracy, and exclude redundant age-metallicity combinations and complex star formation histories from our templates. The final template set we use consists of simple stellar populations with fixed solar metallicity and a \citet{Salpeter55} initial mass function, but spanning a wide range in ages from $1\;$Myr to $15\;$Gyr. We have verified that the quantities derived here are not affected by this choice, while the gain in computing time is significant.

\begin{figure*}
\includegraphics[width=0.49\textwidth]{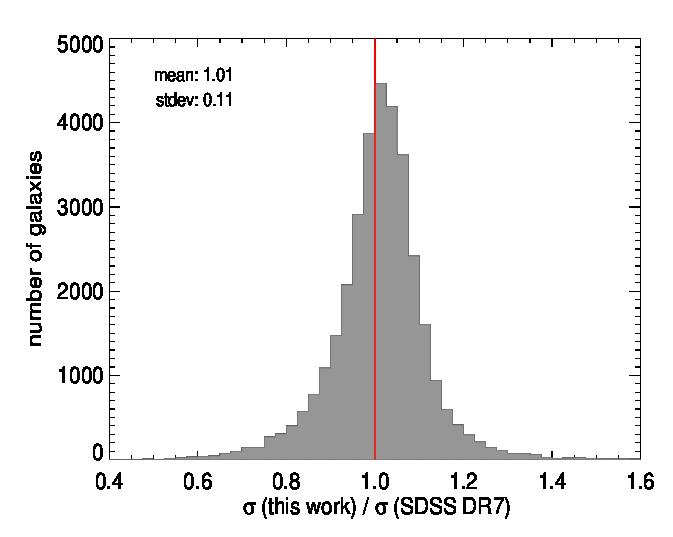}
\includegraphics[width=0.49\textwidth]{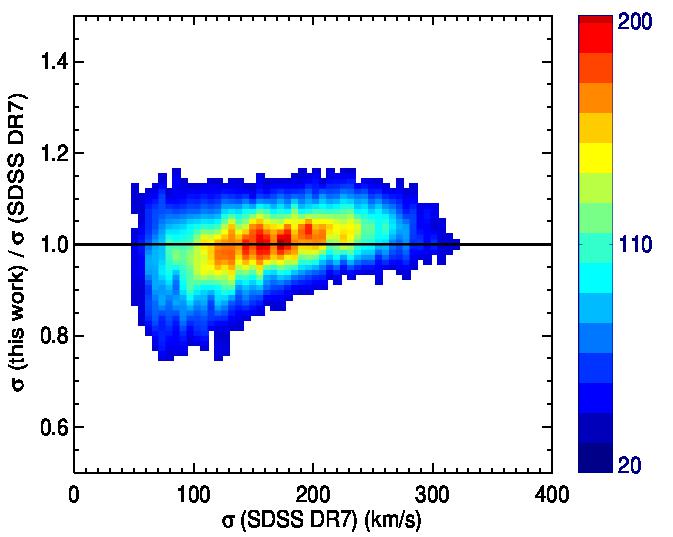}
\caption{Velocity dispersions measured in this work for a subset of SDSS galaxies in comparison with the measurements published in the Data Release 7. {\bf\rm\rm Colour indicates number of galaxies (scale given by the colour bar on the right-hand side)}. There is good agreement between the measurements. The median offset is 1 per cent with a dispersion of 11 per cent.}
\label{fig:sigcomp}
\end{figure*}
\subsection{Stellar velocity dispersion}
Stellar kinematics are evaluated by pPXF. The line-of-sight velocity distribution is fitted directly in pixel space, which is alternative to but consistent with Fourier methods \citep[e.g.][]{Bender90a,RixWhite92}.

A precise knowledge of the spectral resolution of the underlying stellar library is essential for the accurate derivation of stellar velocity dispersion. For this reason, \citet{Beifiori11} have performed a careful analysis of the spectral resolution of the \citet{Mastro11} stellar population models. They have determined resolutions for stellar spectra from the MILES library as well as the \citet{Mastro11} stellar population models through template fits with independent, high-resolution empirical and theoretical spectra. The spectral resolution of the MILES library corresponds to a wavelength independent dispersion $\Delta\lambda$ of $2.54\;$\AA\ (FWHM). This is slightly larger than the $\Delta\lambda=2.3\;$\AA\ indicated in \citet{Sanchez06a} resulting in a lower spectral resolution. This new resolution value has also been found by \citet{Falcon11} and \citet{Prugniel11} in independent re-analyses of the MILES library.

We have used the resolution measurement from \citet{Beifiori11} to downgrade the stellar population templates to the BOSS resolution. {\bf\rm\rm Note that a $\Delta\lambda$ of $2.54\;$\AA\ corresponds to a resolution of $R\sim 1,800$ at $4,500\;$\AA\ (or $\sim 71\;$km$\;$s$^{-1}$), which is the nominal SDSS and BOSS resolution at these wavelengths. As a result, the \citet{Mastro11} templates only require a modest downgrading at longer wavelengths to the correct BOSS resolution.}

At wavelengths blueward $\sim 4,500\;$\AA\, our MILES-based templates have slightly worse spectral resolutions than our data, which can affect the derivation of stellar velocity dispersion. However, we only use the rest-frame wavelength range $4,500-6,500\;$\AA\ for the derivation of stellar velocity dispersion, which is the wavelength range most suitable for stellar kinematics analyses because absorption features are most pronounced \citep{Bender90a,BSG94}. Therefore, the loss of spectral resolution of the MILES library relative to the BOSS resolution at $\lambda\la 4,500\;$\AA\ is not a concern for the present work.

We found good agreement between our measurements of stellar velocity dispersion and the ones published in DR7. The measurements in DR7 are based on the direct fitting method, the same approach adopted by pPXF. Fig.~\ref{fig:sigcomp} shows this comparison. The velocity dispersions are in good agreement with a negligible median offset in $\sigma$ of 1 per cent with a dispersion of 11 per cent.

\subsection{Emission line measurements}
\begin{figure*}
\includegraphics[width=0.35\linewidth]{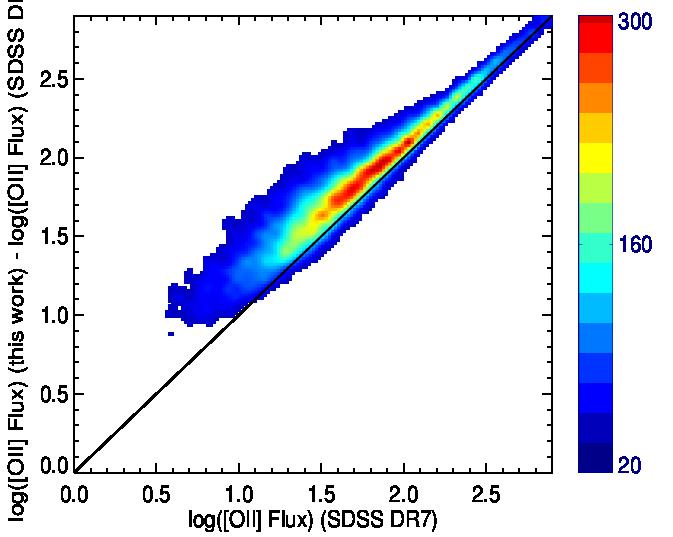} 
\includegraphics[width=0.35\linewidth]{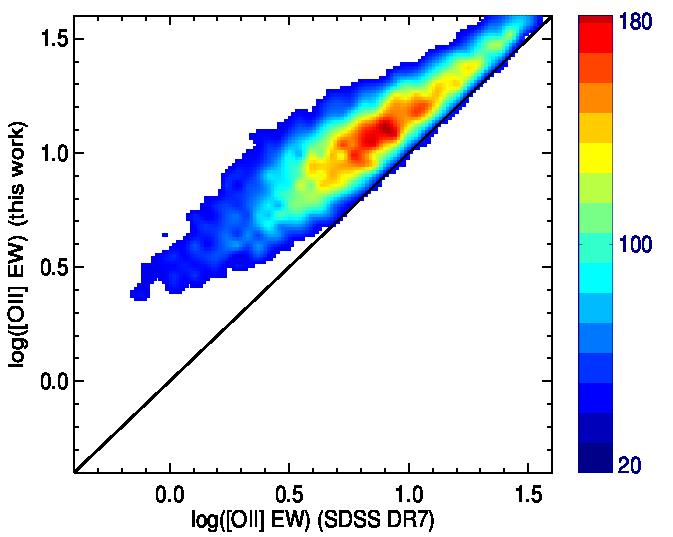}
\includegraphics[width=0.35\linewidth]{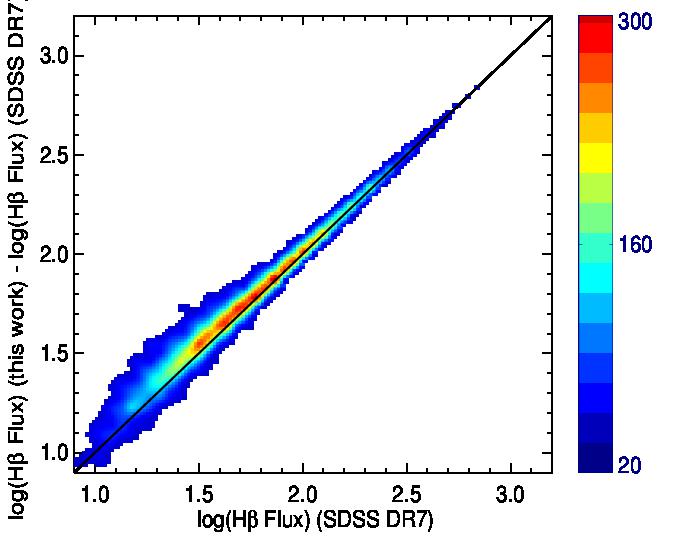} 
\includegraphics[width=0.35\linewidth]{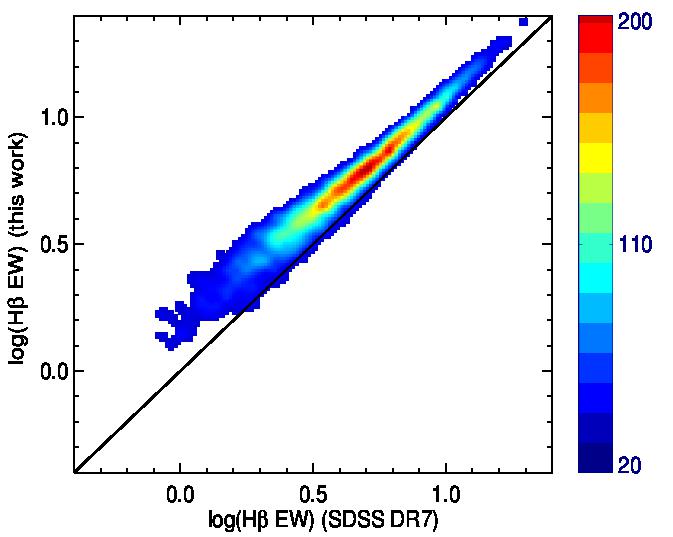}
\includegraphics[width=0.35\linewidth]{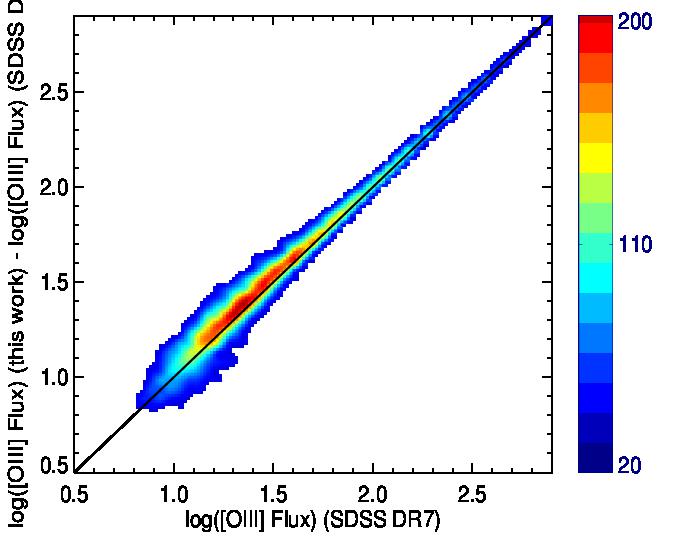} 
\includegraphics[width=0.35\linewidth]{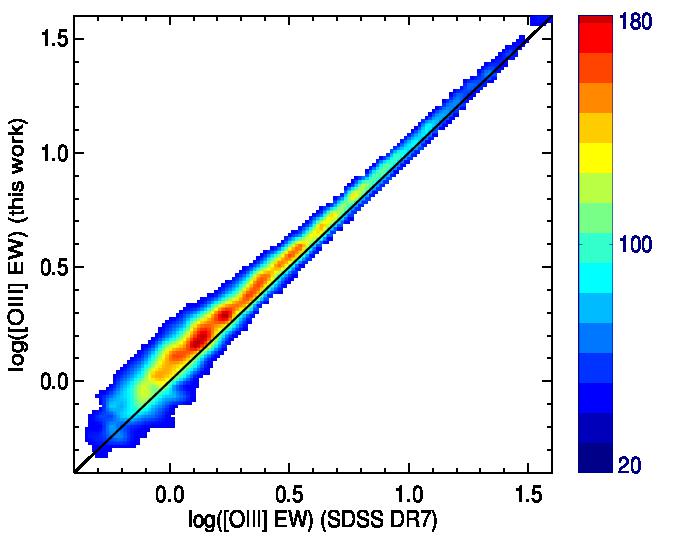}
\includegraphics[width=0.35\linewidth]{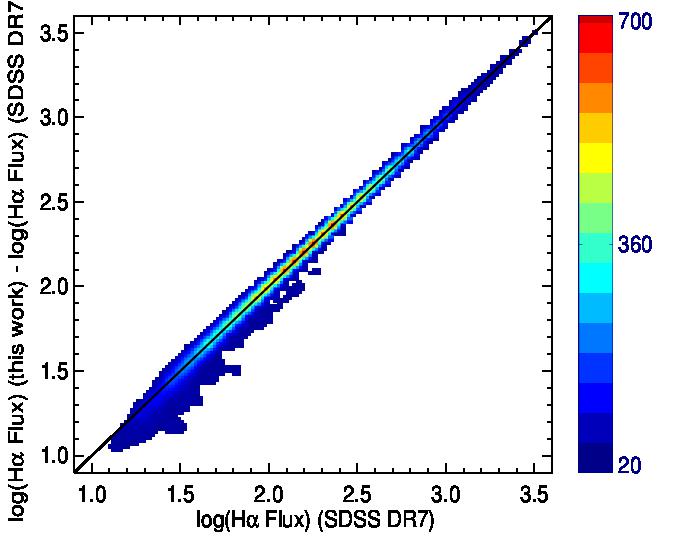} 
\includegraphics[width=0.35\linewidth]{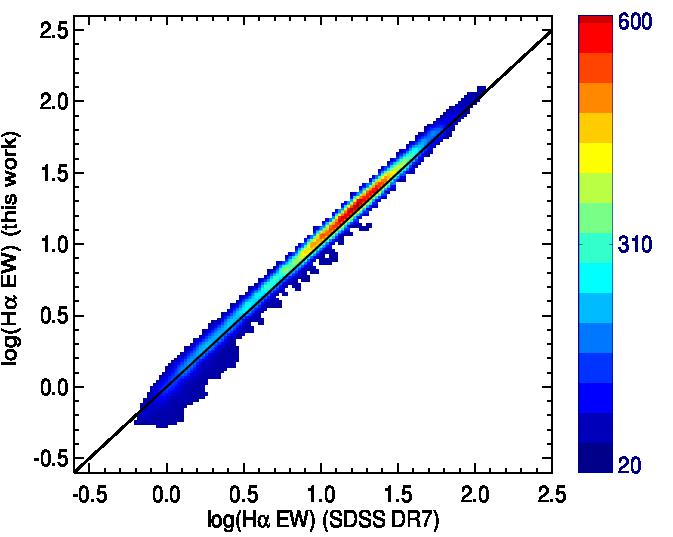}
\includegraphics[width=0.35\linewidth]{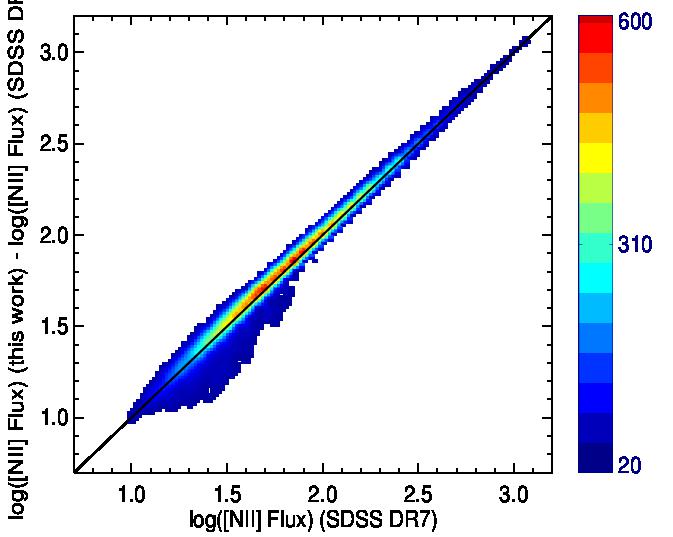} 
\includegraphics[width=0.35\linewidth]{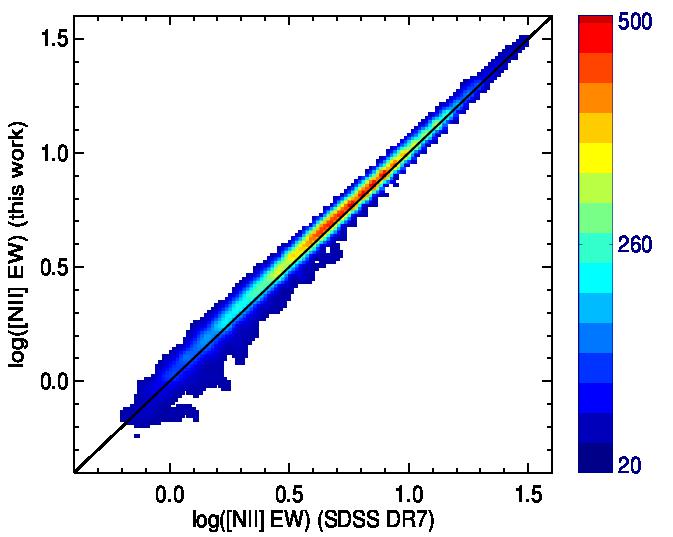}
\caption{Emission line fluxes and equivalent widths for {\bf\rm [OII]$\lambda\lambda 3726+3729$}, \Hb, [OIII]$\lambda 5007$, \Ha, and [NII]$\lambda 6583$ measured in this work for a subset of SDSS-I/II galaxies in comparison with the measurements published in the Data Release 7. {\bf\rm [OII]$\lambda\lambda 3726+3729$ has been calculated as the sum of [OII]$\lambda 3726$ and [OII]$\lambda 3729$.} {\bf\rm Colour indicates number of galaxies (scale given by the colour bar on the right-hand side)}. Observed, non dust-corrected, values are used. Emission line fluxes and equivalent widths are slightly higher in this work by $\sim 0.1\;$dex.}
\label{fig:emcomp}
\end{figure*}
\begin{figure*}
\includegraphics[width=\textwidth]{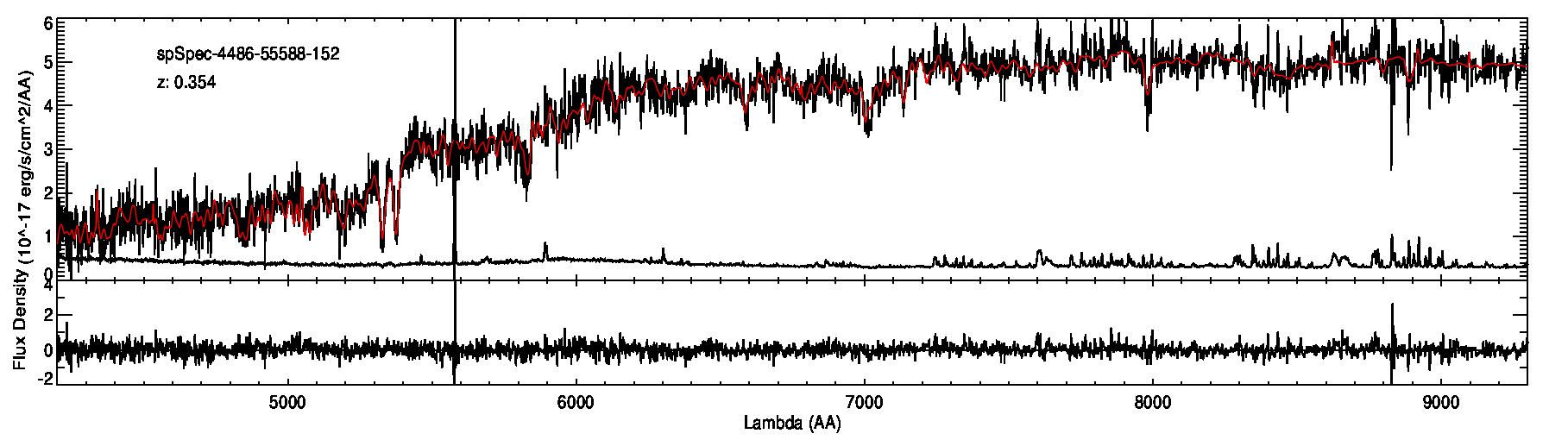}
\includegraphics[width=\textwidth]{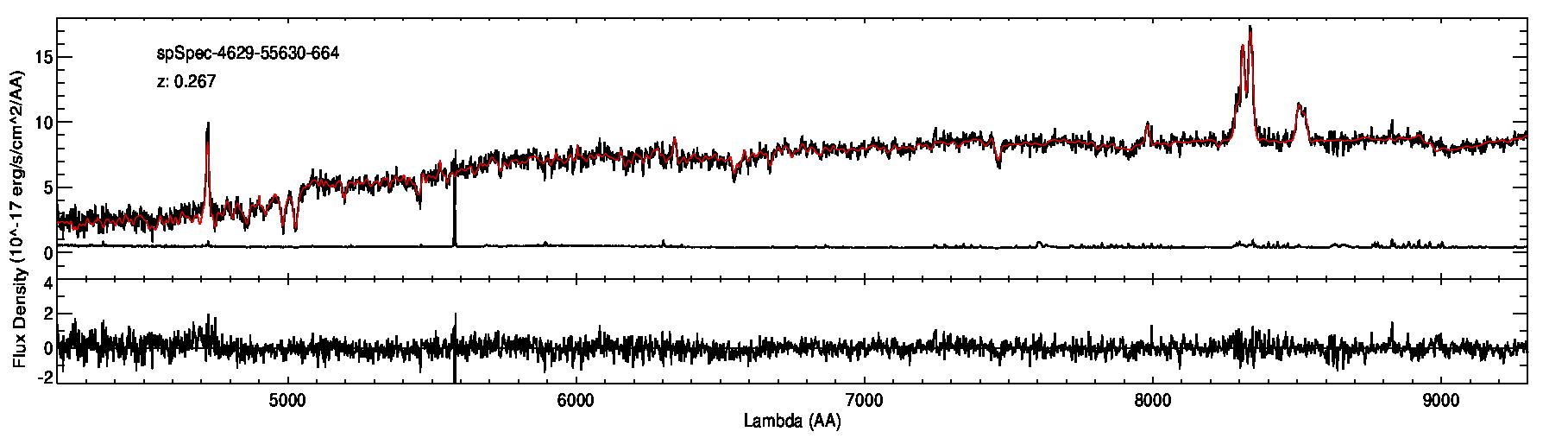}
\caption{Example spectra with and without emission lines for BOSS galaxies around $z\sim 0.3$ in observed-frame wavelength with the best fit spectrum composed of the stellar population and the emission line templates overplotted (red line). The bottom spectrum is the variance, the residual from the fit is plotted in the bottom sub-panel.}
\label{fig:spectra}
\end{figure*}
\begin{figure*}
\includegraphics[width=0.33\textwidth]{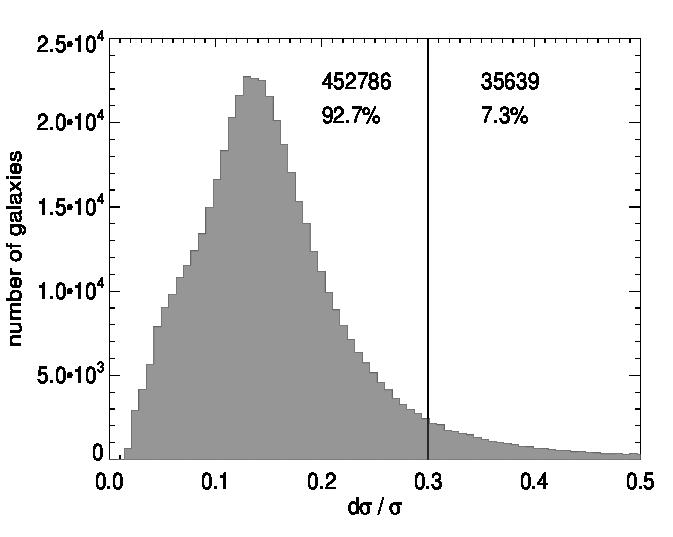}
\includegraphics[width=0.33\textwidth]{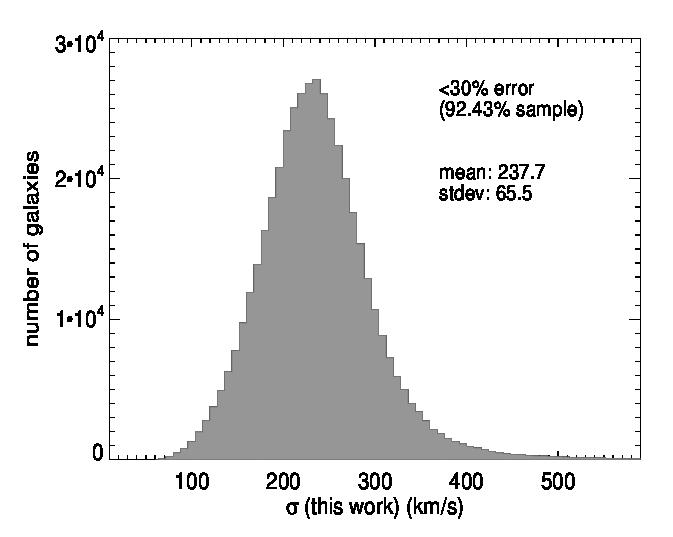}\\
\includegraphics[width=0.33\textwidth]{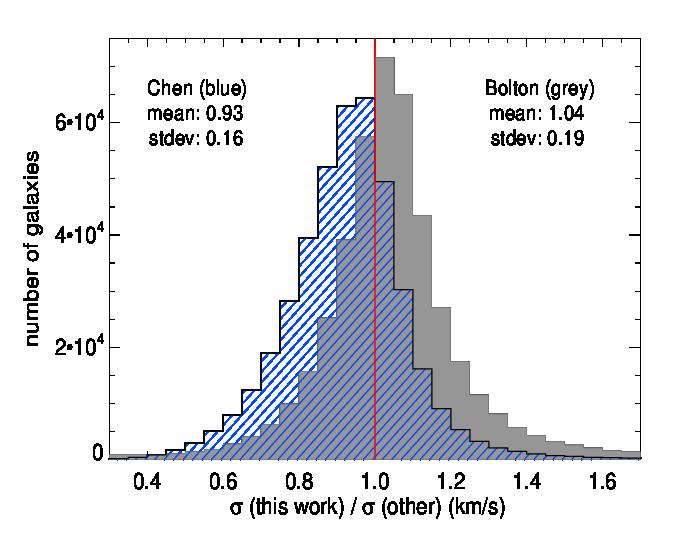}
\includegraphics[width=0.33\textwidth]{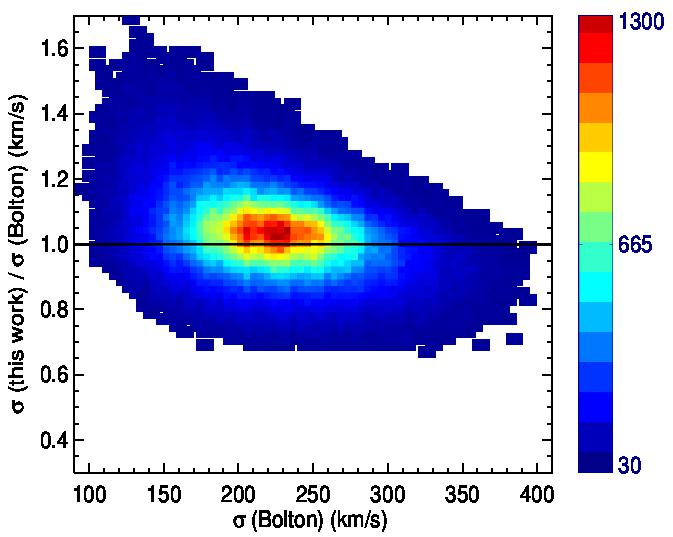}
\includegraphics[width=0.33\textwidth]{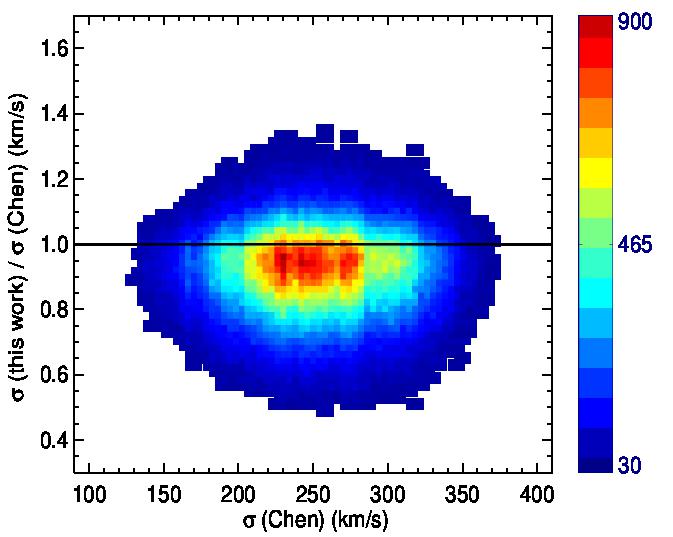}
\caption{{\em Top panels:} Distributions of the relative errors in the velocity dispersion measurements of BOSS galaxies  and the distribution of velocity dispersions with error smaller than 30 per cent. {\em Bottom panels:} Comparison with independent measurements of velocity dispersions based on the original SDSS DR7 pipeline code by \citet{Bolton12} and the code by \citet{Chen12}. {\bf\rm Colour indicates number of galaxies (scale given by the colour bar on the right-hand side). There are small systematic offsets. The measurements of this work are $\sim 4$ per cent higher than in \citet{Bolton12} and  $\sim 7$ per cent lower than in \citet{Chen12} at dispersions of 19 and 16 per cent, respectively.}}
\label{fig:sigmacomp}
\end{figure*}

{\bf\rm Rest-frame} emission line fluxes and EWs of all emission lines available in the rest-frame wavelength range covered are measured on the emission line template spectra. Table~\ref{tab:elines} summarises the list of emission lines measured, if accessible in the rest-frame spectrum. 
An emission line measurement is made when the amplitude-over-noise (AoN) ratio is larger than two, which implies that many of these lines are too weak to be detected in BOSS. Reasonable AoN ratios can be obtained in a subsample of BOSS galaxies, however, for some of the stronger lines such as the forbidden lines [OII]$\lambda\lambda 3726+3729$, [OIII]$\lambda 5007$, and [NII]$\lambda 6583$, as well as the Balmer lines \Hb\ and \Ha. {\bf\rm Line fluxes of the OII doublet [OII]$\lambda\lambda 3726+3729$ are calculated as the sum of the individual lines [OII]$\lambda 3726$ and [OII]$\lambda 3729$.} EWs are calculated as the ratio between line and continuum fluxes. The continuum flux adopted is the median of the average fluxes in the two blue and red continuum windows at a distance of $200\;$km$\;$s$^{-1}$ from the centre of the line and with a width of $200\;$km$\;$s$^{-1}$.

{\bf\rm We note that continuum fluxes as listed in the Portsmouth DR9 catalogue still need to be corrected to rest-frame by multiplication with the factor $(1+z)$. Consequently, EW provided in the Portsmouth DR9 catalogue need to be divided by the same factor $(1+z)$ to be translated into rest-frame. To be better consistent with the emission line fluxes in the catalogue, that are presented in rest-frame, future data releases (DR10 and onwards) will provide proper rest-frame continuum fluxes and EWs with this correction factor already included.}

\begin{table}
\caption{Emission lines measured.}
\begin{center}
\begin{tabular}{cccc}
\hline\hline
Index & Name & $\lambda$ (\AA) & free/tied\\
\hline
1 & HeII & 3203.15 & free\\
2 & [NeV] & 3345.81 & free\\
3 & [NeV] & 3425.81 & free\\
4 & [OII] & 3726.03 & free\\
5 & [OII] & 3728.73 & free\\
6 & [NeIII] & 3868.69 & free\\
7 & [NeIII] & 3967.40 & free\\
8 & H5 & 3889.05 & free\\
9 & H$_{\epsilon}$ & 3970.07 & free\\
10 & H$_{\delta}$ & 4101.73 & free\\
11 & H$_{\gamma}$ & 4340.46 & free\\
12 & [OIII] & 4363.15 & free\\
13 & HeII & 4685.74 & free\\
14 & [ArIV] & 4711.30 & free\\
15 & [ArIV] & 4740.10 & free\\
16 & H$_{\beta}$ & 4861.32 & free\\
17 & [OIII] & 4958.83 & tied to 18\\
18 & [OIII] & 5006.77 & free\\
19 & [NI] & 5197.90 & free\\
20 & [NI] & 5200.39 & free\\
21 & HeI & 5875.60 & free\\
22 & [OI] & 6300.20 & free\\
23 & [OI] & 6363.67 & free\\
24 & [NII] & 6547.96 & tied to 26\\
25 & H$_{\alpha}$ & 6562.80 & free\\
26 & [NII] & 6583.34 & free\\
27 & [SII] & 6716.31 & free\\
28 & [SII] & 6730.68 & free\\
\hline
\end{tabular}
\end{center}
\label{tab:elines}
\end{table}

In case of a relatively low signal-to-noise ratio as provided by SDSS and BOSS spectroscopy, it would generally be preferable to tie the kinematics of most emission lines to stronger related lines such as \Ha\ at $6563\;$\AA\ and [NII] at $6583\;$\AA\ \citep{Brinchmann04,Sarzi06,Schawinski07b}. This is not possible for BOSS, however, because of the relatively large redshift range probed. These key emission lines cannot be observed with an optical CCD for objects beyond redshift $z\sim 0.5$. Velocities and velocity dispersions are therefore free parameters for all emission lines, except for [OIII]$\lambda 4959$ and [NII]$\lambda 6548$ that are tied to the stronger nearby lines [OIII]$\lambda 5007$ and [NII]$\lambda 6583$, respectively (see Column 4 in Table~\ref{tab:elines}). We follow this approach independently of redshift to guarantee homogeneity within the full sample.

The \textsc{GANDALF} code considers two dust components: the diffuse dust in the galaxy affecting the spectral shape and dust in emission line regions additionally affecting emission line fluxes and ratios. The latter is estimated through the Balmer decrement between \Hb\ and \Ha\ when available \citep{VeilleuxOsterbrock87}. However,  the relatively low signal-to-noise ratios in the BOSS spectra make the measurement of the Balmer decrement highly uncertain. Cases in which the H$\beta$ emission line is barely detected, the inclusion of this second dust component tends to yield unreasonably high values for dust extinction. Thus, we do not consider this second dust component in the fits and focus on the diffuse dust component that only affects the spectral shape adopting a \citet{Calzetti01} obscuration curve. The emission line fluxes provided have been corrected for dust extinction obtained in this way. Note that we do not correct for Milky Way foreground extinction before the emission lines analysis. This does not affect the emission line measurements, but does imply that the resulting E(B-V) values need to be corrected for Milky Way extinction a posteriori.

As mentioned at the beginning of Section~\ref{sec:tool}, we compare our emission line measurements of a subset of SDSS DR7 galaxies with the MPA-JHU values published in DR8 \citep{Brinchmann04,Tremonti04}. Observed, not dust-corrected, values are used. {\bf\rm As MPA-JHU emission line fluxes are corrected for Milky Way foreground dust reddening, we apply this correction to our uncorrected fluxes for this comparison. To this end we extracted the E($B-V$) values for each object from the MPA-JHU catalogue \citep{Schlegel98}, and calculated the attenuation for each line utilising the \citet{ODonnell94} extinction curve. Furthermore, it should be noted that MPA-JHU line fluxes are rescaled on a per-plate basis such that the mean $r$-band flux in the spectrum matches the $r$-band fiber mag from the photometry. We have therefore multiplied our line fluxes with this rescaling factor provided by the "spectofiber" keyword in the MPA-JHU database. Finally, as explained above, we divide our DR9 EW measurements by the factor $(1+z)$ to ensure EWs are measured in rest-frame.}

{\bf\rm Fig.~\ref{fig:emcomp} shows the resulting comparison for emission line fluxes and EWs of [OII]$\lambda\lambda 3726+3729$, \Hb, [OIII]$\lambda 5007$, \Ha, and [NII]$\lambda 6583$. It can be seen that the measurements generally agree well showing tight correlations with small scatter and only small offsets. Median offsets in emission line flux measurements are below $0.02\;$dex with a dispersion of $\sim 0.1\;$dex. Only [OII]$\lambda\lambda 3726+3729$ is slightly more offset by $\sim 0.1\;$dex with a somewhat larger dispersion of $\sim 0.2\;$dex. This may not be surprising, as the [OII]$\lambda\lambda 3726+3729$ doublet is barely resolved at SDSS spectral resolution, and the measurement is therefore more uncertain. Similar offsets, even though somewhat larger, are present for the EWs. Median offsets are below $0.04\;$dex for [OIII]$\lambda 5007$, \Ha, and [NII]$\lambda 6583$, while they increase to $0.1\;$dex and $0.25\;$dex for \Hb\ and [OII]$\lambda\lambda 3726+3729$, respectively, with a larger dispersion of $0.2\;$dex. These larger discrepancies in the EW measurements will most likely be caused by differences in the treatment of continuum fitting. As for the fluxes, discrepancies in [OII]$\lambda\lambda 3726+3729$ are further caused by uncertainties in the measurement of this line, as the doublet is not resolved.}

We note that this same version of the code is being used within the Galaxy and Mass Assembly (GAMA) project \citep{Driver11} to derive simple stellar kinematics and emission line fluxes. The GAMA spectroscopy and a comparison of GANDALF-based emission line measurements with other independent methods are presented in \citet{Hopkins12}. In this comparison, particular care has been taken in the homogenisation of the treatment of dust reddening. The result reinforces the above conclusion. It is shown that the measurements of emission line fluxes are consistent to better than 5 per cent, with a dispersion consistent with the error measurements on the lines. The line EW estimates are also consistent to better than 5 per cent in the red, and to $10-15$ per cent in the blue, again with dispersions consistent with the measured errors.

\section{Velocity dispersions of BOSS galaxies}
\label{sec:velocity}
In this section we present the measurements of the line-of-sight velocity dispersions for the BOSS galaxy sample. We show their error distributions, compare with independent measurements and discuss their redshift evolution.

\subsection{Example spectra in BOSS}
{\bf\rm Fig.~\ref{fig:spectra} shows two example spectra for BOSS galaxies around $z\sim 0.3$ with the best fit spectrum composed of the stellar population and the emission line templates overplotted. It can be seen that the code is successful at identifying true emission line features over the relatively high noise level. The stellar population fits are good enough to make measurement of stellar kinematics and emission line fluxes. Results and reliability will be assessed in the following sections. As mentioned earlier, a more comprehensive approach with a wider set of templates will be required to derive reliable star formation histories from the spectra. Note, however, that accurate derivation of stellar population parameters from individual BOSS spectra is challenging because of the modest S/N ratios, and the stacking of spectra to obtain better S/N ratios is advisable \citep{Chen12}.}

\begin{figure*}
\includegraphics[width=0.49\textwidth]{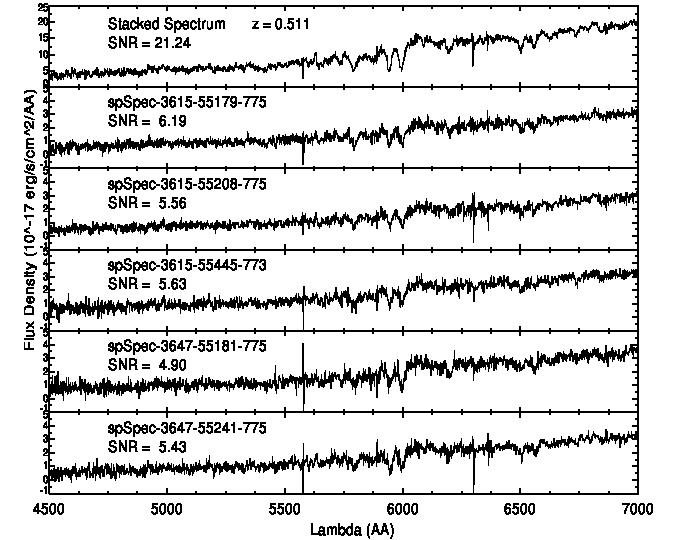}
\includegraphics[width=0.49\textwidth]{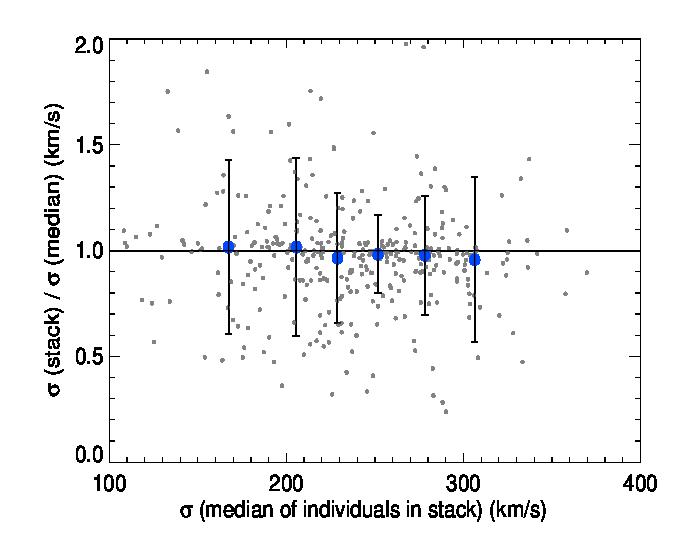}
\caption{{\em Left-hand panel:} Individual spectra and their stack of a single BOSS galaxy with repeated 1-hour observations. The S/N ratio per resolution element is given by the labels. The S/N ratio of the stacked spectrum is higher by about a factor four compared to the individual spectra. {\em Right-hand panel:} Median velocity dispersion measured on individual BOSS spectra versus the {\bf\rm ratio} between the median of measurements on individual and stacked spectra. Each stacked spectrum is the sum of individual spectra of the same object from repeated plate observations. The blue symbols are the median in bins of velocity dispersion, {\bf\rm error bars indicate Poisson errors}. The agreement is very good, hence velocity dispersion measurements are reliable even at typical BOSS S/N.}
\label{fig:stackedspectrum}
\end{figure*}
\subsection{Velocity dispersions and their errors}
The typical signal-to-noise ratio in each BOSS galaxy spectrum is about $5\;$\AA$^{-1}$. This is considerably lower than the average provided by SDSS-I spectroscopy owing to the faintness of the BOSS targets. However, despite this limitation, measurements of stellar velocity dispersion at reasonable accuracy can still be made. The top-left panel in Fig.~\ref{fig:sigmacomp} shows the distribution in formal measurement errors. There is a strong peak at a typical relative error of 14 per cent, and 93 per cent of the objects have a relative error below 30 per cent. {\bf\rm Hence, errors are relatively large, as expected from the low signal-to-noise ratios, but $\sigma$ measurements are robust as discussed in the following section through comparison with independent measurements.} The top-right panel in Fig.~\ref{fig:sigmacomp} shows the distribution of velocity dispersions with error smaller than 30 per cent. The distribution is approximately Gaussian with a well pronounced peak at $\sigma=240\;$km/s. This relatively high value is consistent with the goal of the BOSS galaxy target selection algorithm designed to target massive galaxies \citep{Padmanabhan10,Dawson12,Maraston12}.

\subsection{Comparison with independent measurements}
To test the reliability of these measurements we compare our velocity dispersions with alternative measurements. In the bottom panels of Fig.~\ref{fig:sigmacomp} we compare our results with two independent measurements, one from \citet{Bolton12} based on the original SDSS pipeline \citep[used in][]{Shu11} and the other from \citet{Chen12}. {\bf\rm The velocity dispersions are slightly higher by $\sim 4$ per cent than the ones of \citet{Bolton12} at a dispersion of $5$ per cent and slightly smaller by $\sim 7$ per cent than the ones of \citet{Chen12} at dispersions of 19 and 16 per cent, respectively.} Both these offsets are {\bf\rm smaller than the typical measurement error of 14 per cent}. This result is encouraging and underlines the reliability of stellar velocity dispersion measurements on the BOSS spectra despite the relatively large error and low signal-to-noise ratio.

\begin{figure*}
\includegraphics[width=0.32\textwidth]{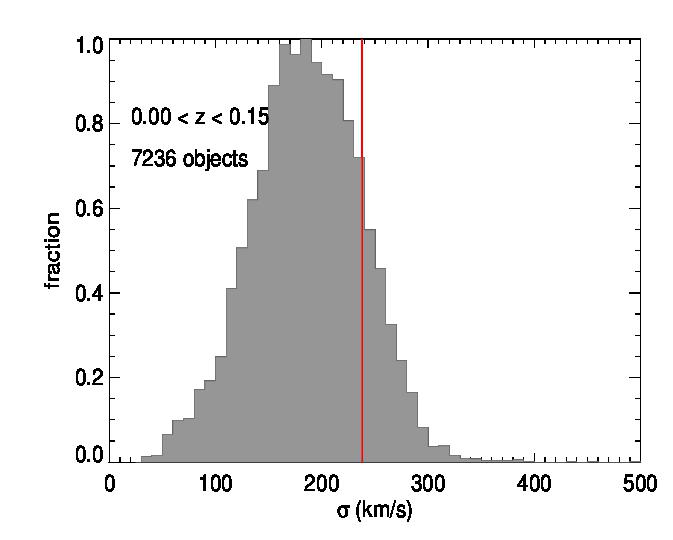}
\includegraphics[width=0.32\textwidth]{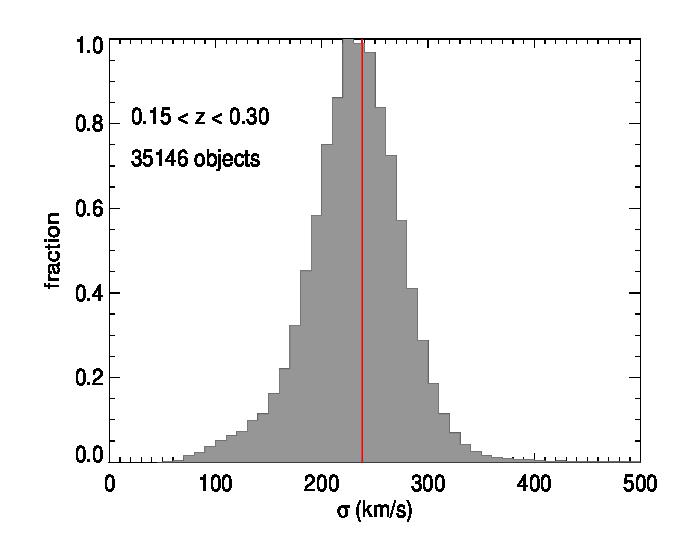}
\includegraphics[width=0.32\textwidth]{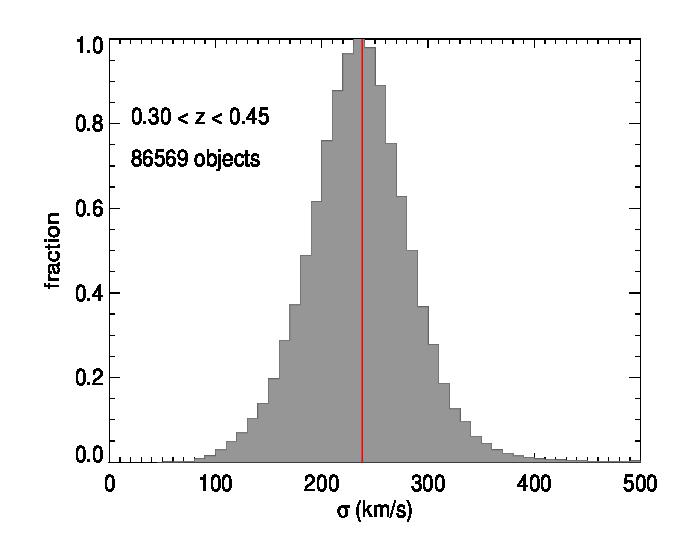}
\includegraphics[width=0.32\textwidth]{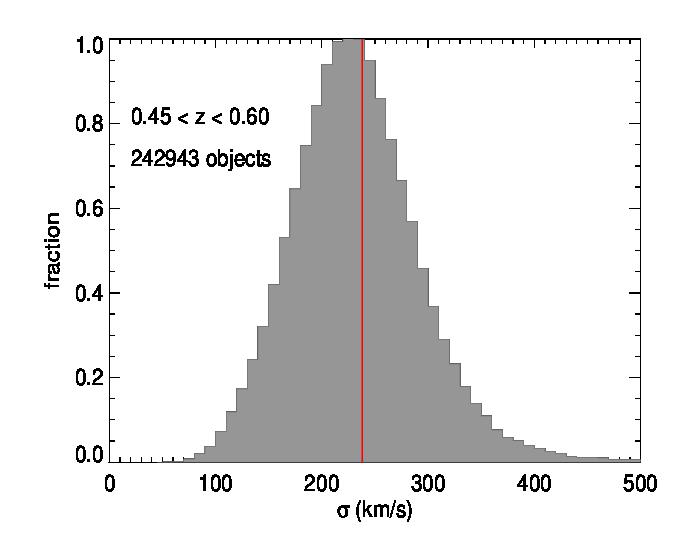}
\includegraphics[width=0.32\textwidth]{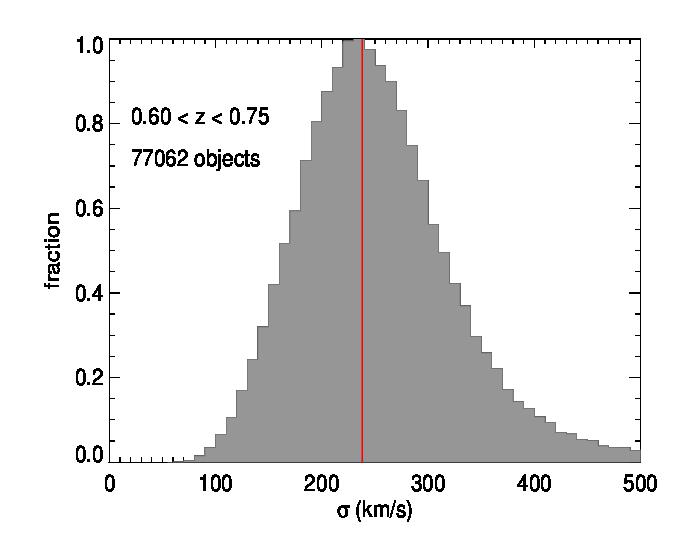}
\includegraphics[width=0.32\textwidth]{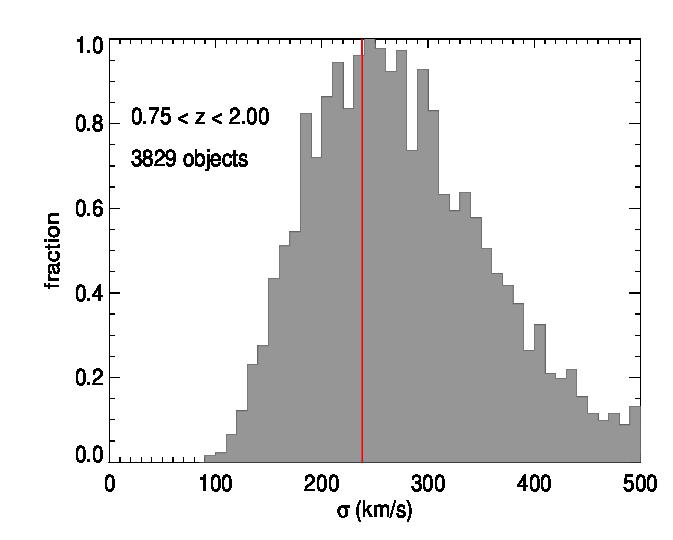}
\caption{Distribution in velocity dispersion for spectra with an error of less than 30 per cent in the redshift intervals $0< z\leq 0.15$, $0.15< z\leq 0.3$, $0.3< z\leq 0.45$, $0.45< z\leq 0.6$, $0.6< z\leq 0.75$, and $z>0.75$. The red line indicates the position of the peak for the four central redshift intervals. The distribution is approximately Gaussian and there is no evolution within $0.15<z<0.75$. At redshifts below $z\sim 0.15$ and above $z\sim 0.75$, the distributions are skewed and slightly offset toward lower and higher velocity dispersions, respectively.}
\label{fig:sigmared}
\end{figure*}

\subsection{Comparison with higher S/N spectra}
Additionally to the comparison with independent measurements we investigated the dependence on S/N ratio. To this end we made use of repeated plate observations in the course of the last two years. We constructed a sample of 574 BOSS galaxies mainly from plates 3615 and 3647 with at least six 1-hour observations each. We then stacked the spectra from the individual exposures to a high S/N spectrum for each object. The left-hand panel of Fig.~\ref{fig:stackedspectrum} shows the individual spectra and the final stack. It can be seen that the resulting spectrum has a significantly higher S/N ratio. The latter increases by about a factor 4 from $\sim 5$ to $\sim 20$ per resolution element. It is important to note that each stacked spectrum represents the stack for one individual object.

Finally, we measured stellar velocity dispersion on both the individual and the stacked spectra. The result of this exercise is presented in the right-hand panel of Fig.~\ref{fig:stackedspectrum}, where we plot the {\bf\rm ratio} between the measurements on individual and stacked spectra as a function of the measurements on individual spectra. The agreement is very good. There is no systematic offset between measurements on individual (low S/N) and stacked (high S/N) spectra. Furthermore, we find no evidence for a systematic trend with S/N ratio. Hence, overall this exercise demonstrates that velocity dispersion measurements on BOSS spectra are reliable (within their errors) in spite of the relatively low S/N ratios.

\subsection{Distribution of velocity dispersions as a function of redshift}
In Fig.~\ref{fig:sigmared} we show the distributions of velocity dispersions with an error less than 30 per cent, corresponding to 93 per cent of the sample in the redshift intervals $0-0.15$, $0.15-0.3$, $0.3-0.45$, $0.45-0.6$, $0.6-0.75$, and for $z>0.7$. The red line indicates the mean position of the peak for the four central redshift intervals. 

Fig.~\ref{fig:sigmared} shows that the distribution in velocity dispersion is largely independent of redshift within $0.15<z<0.75$. The position of the peak of the distribution does not change, while the width becomes somewhat larger at higher redshifts caused by slightly larger measurement errors \citep[see also][]{Shu11}. At redshifts below $z\sim 0.15$ and above $z\sim 0.75$, the distributions are skewed and slightly offset toward lower and higher velocity dispersions, respectively. This uniformity of the distribution in velocity dispersion as a function of redshift between 0.15 and 0.75 is an important feature of the BOSS target selection, as it establishes a direct link between the galaxy samples at various redshifts, which allows us to accurately probe the redshift evolution of massive galaxies. Target selection in BOSS has been designed to this purpose and, using stellar velocity as proxy for galaxy mass, the present results shows that this goal has been achieved. A similar conclusion is drawn in \citet{Maraston12}, who show that the distribution of photometric stellar masses is approximately uniform up to $z\sim 0.6$. In \citet{Beifiori12} we use these velocity dispersion measurements for the derivation of dynamical galaxy masses and discuss the evolution of the dynamical-to-stellar mass ratios of BOSS galaxies with redshift.

\section{Emission line properties of BOSS galaxies}
\label{sec:emission}
In this section we present the emission line properties of our BOSS galaxy sample. The aim is to discuss the prevalence of various galaxy types and emission line classes in the BOSS target selection algorithm. {\bf\rm We split the analysis in two parts. In Section~\ref{sec:bpt} we focus on} galaxies with redshifts $z<0.45$, so that the full set of key diagnostic emission lines \Hb, [OIII]$\lambda 5007$, \Ha, and [NII]$\lambda 6583$ can be measured in the BOSS spectrum. This sample contains 140,596 galaxy spectra out of the 492,450 analysed in the full redshift range. Note that in this way we deselect a large fraction of CMASS galaxies, the redshift distribution of which peaks at $z\sim 0.6$. {\bf\rm Therefore, we present an additional analysis of emission line properties of CMASS galaxies at $z>0.3$ in Section~\ref{sec:lama} using alternative but more restrictive methods based on bluer emission lines within the observed wavelength range. We only include objects where the full set of key diagnostic emission lines can be detected with an AoN above 2 to allow for a proper analysis of the emission line characteristics through emission line ratio diagnostic diagrams.}

\begin{figure*}
\includegraphics[width=0.33\textwidth]{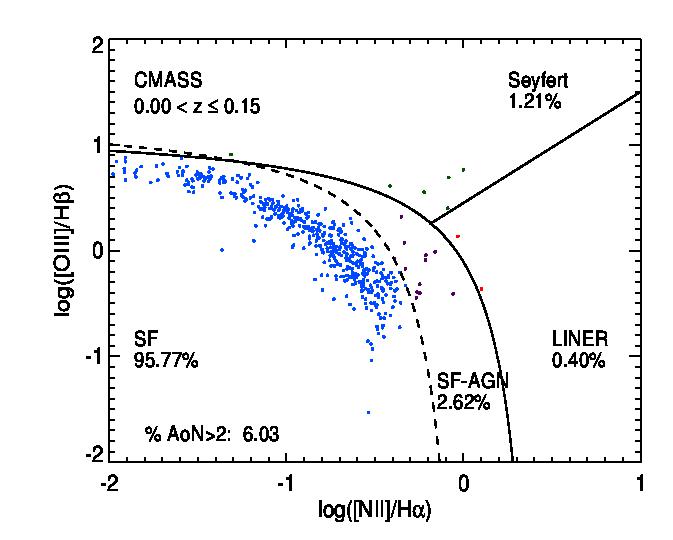}
\includegraphics[width=0.33\textwidth]{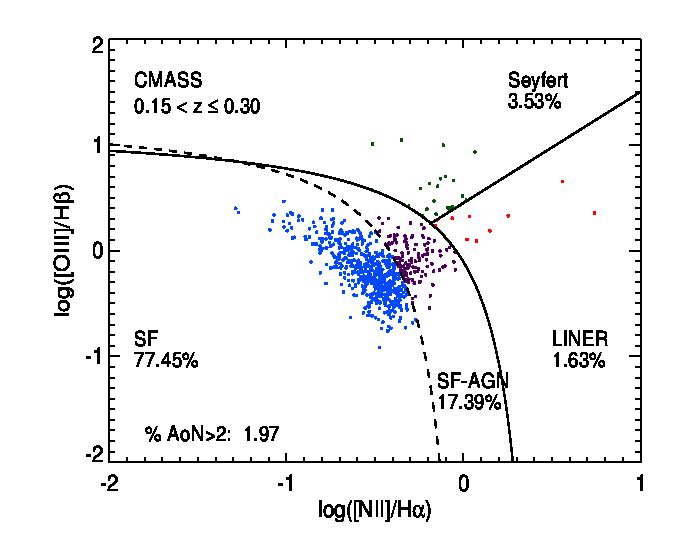}
\includegraphics[width=0.33\textwidth]{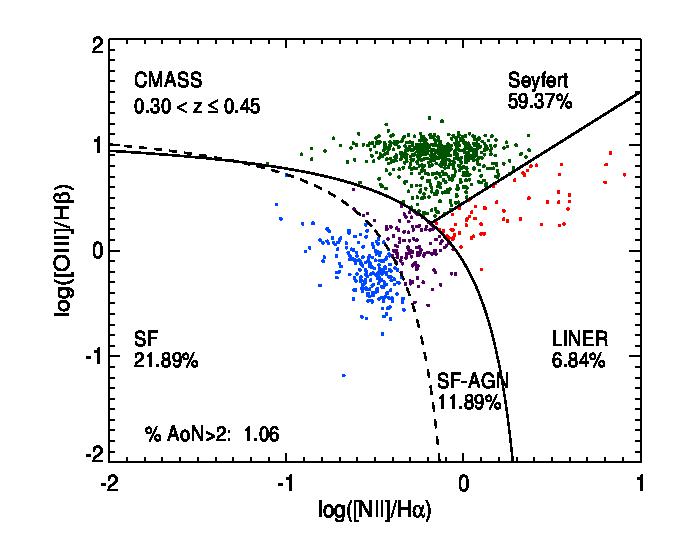}
\includegraphics[width=0.33\textwidth]{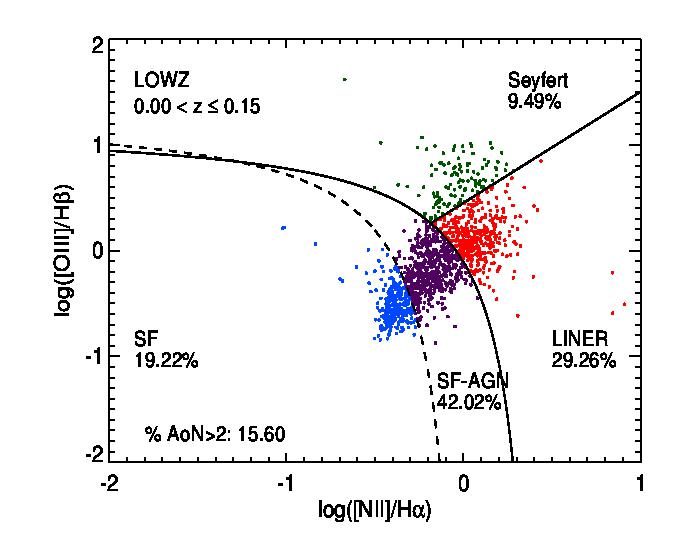}
\includegraphics[width=0.33\textwidth]{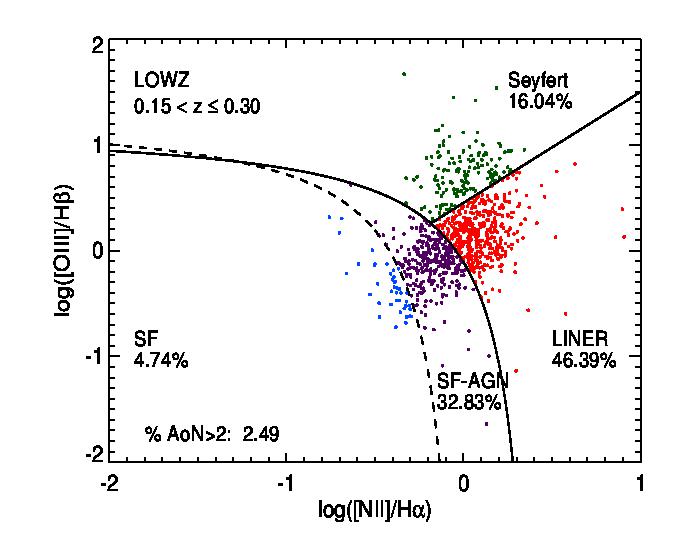}
\includegraphics[width=0.33\textwidth]{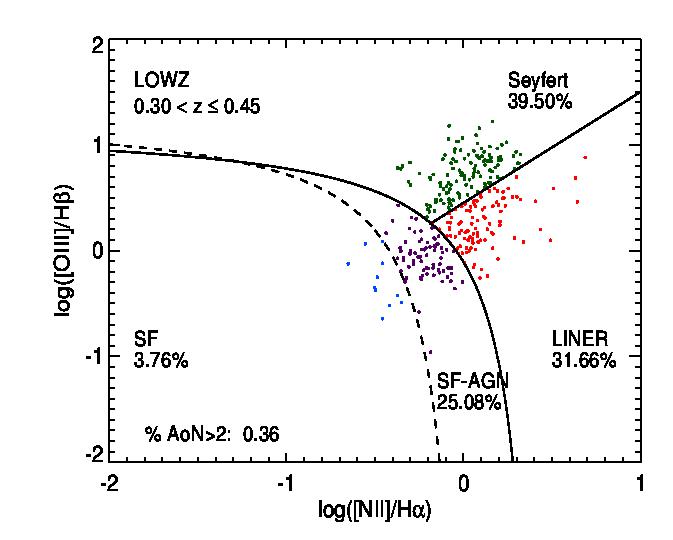}
\caption{Emission line classification \citep{BPT81} for BOSS galaxies at various redshift bins (redshift increasing from left to right) for CMASS galaxies (top panels) and LOWZ galaxies (bottom panels). Objects at $z<0.45$ are selected to warrant detectability of \Ha\ and [NII]$\lambda 6583$. The fraction of objects for which all four emission lines are detected (requiring the amplitude-over-noise ratio of all four lines to be larger than two) is given in the bottom-left corner of each panel. The empirical separation between star forming galaxies and AGN (dashed line) is from \citet{Kauffmann03b}, and the theoretical extreme star burst line from \citet{Kewley01} to identify pure AGN emission (solid curved line). A dividing line defined by \citet{Schawinski07b} is used to distinguish between LINER and Seyfert emission (solid straight line). The fractions of galaxies in each of the emission line classes are given by the labels in the panels. Overall, the fraction of objects with detected emission lines is small in BOSS. There is a marked difference in the emission line properties between the BOSS galaxy samples at high and low redshift, mainly caused by selection effects. Furthermore, there is a striking difference between the LOWZ and CMASS samples (see text for details).}
\label{fig:bpt}
\end{figure*}

\subsection{BPT classification}
\label{sec:bpt}
We use the emission lines \Hb, [OIII]$\lambda 5007$, \Ha, and [NII]$\lambda 6583$ to perform the standard classification based on the diagnostic diagram first introduced by \citet{BPT81} and \citet{VeilleuxOsterbrock87}, and widely used in studies of SDSS galaxies \citep[e.g.][]{Kauffmann03b,Miller03,Hopkins03,Brinchmann04,Kewley06,Schawinski07b,Wild07,Stasinska08}. In this method the emission lines ratios [OIII]/\Hb\ and [NII]/\Ha\ are used as indicators for the ionisation source of the interstellar gas in galaxies. The diagram has been proven to be very powerful at separating star forming galaxies from objects with AGN. {\bf\rm The AoN limit of 2 for all four diagnostic emission lines yields a sample of 4,887 spectra (3.5 per cent of the total). Note that the fractions of galaxies with significant detections of the emission lines \Ha\ and [NII]$\lambda 6583$ are considerably higher (around 30 per cent).}

The various panels of Fig.~\ref{fig:bpt} show the BPT classifications for the 4,887 BOSS galaxies in three redshift bins, with redshift increasing from left to right (see labels in the panels). We adopt the empirical separation between star forming galaxies and AGN (dashed line) as defined by \citet{Kauffmann03b}, and the theoretical extreme star burst line by \citet{Kewley01} to identify pure AGN emission (solid curved line). As is commonly done, we assume that the area between these two separating lines is populated by galaxies with a composite of star burst and AGN spectra. We further use the dividing line defined by \citet{Schawinski07b} to distinguish between LINER and Seyfert emission (solid straight line) based on SDSS galaxy classifications obtained through the [SII]/\Ha\ ratio.

It can be seen that, in contrast to the distribution in galaxy masses and stellar velocity dispersion, there is a {\bf\rm change of the emission line properties of BOSS galaxies as a function of redshift}, mainly because of selection effects. Furthermore, there is a striking difference between the LOWZ and CMASS samples. Besides the fact that the number of galaxies selected decreases with increasing redshift for LOWZ and increases for CMASS owing to the design of the target selection, the overall fraction of galaxies with detected emission lines (see label in bottom-left corner in each panel) is different for LOWZ and CMASS and changes with redshift. In general, the fraction of star forming galaxies among BOSS galaxies can be expected to be low, independently of redshift, because of the bias towards massive galaxies at all redshifts.

\subsubsection{LOWZ}
Not surprisingly, the largest fraction of BOSS galaxies with detectable emission lines (15 per cent) is found in the LOWZ sample at redshifts below 0.15. This should be expected as the contamination with lower-mass galaxies is largest (see Fig.~\ref{fig:sigmared}). The majority of emission-line galaxies has some AGN component, the fraction of purely star forming galaxies only being 20 per cent. The fraction of galaxies with detected emission lines drops dramatically with increasing redshift to only a few per cent. The prevalence of AGN and LINER-like emission increases further with increasing redshift above 0.15.

We have also examined the presence of BOSS emission-line galaxies that are likely to be ionised through old stellar populations rather than AGN. Following \citet{CidFernandes11}, we assume that such galaxies can be characterised by having \Ha\ EWs below $3\;$\AA. This threshold has been derived from the distribution of \Ha\ EWs in SDSS galaxies, which is strongly bimodal \citep[see also][]{Bamford08}, with two peaks at 1 and $16\;$\AA\ and an intermediate minimum in the neighbourhood of $3\;$\AA\ \citep{CidFernandes11}. We find that these so-called 'retired' galaxies typically have LINER emission, but only a minority (about one quarter) of galaxies with LINER emission line characteristics is found to be better described by this object class.

\subsubsection{CMASS}
Interestingly, there are some CMASS galaxies at low redshifts below 0.15, even though the selection cuts are designed to target galaxies above $z\sim 0.4$. Fig.~\ref{fig:bpt} shows that most of these objects are star forming galaxies that fell into the CMASS colour selection cut despite being at low redshift, most probably due to dust reddening. These galaxies are also relatively low-mass with velocity dispersions at the low-end of the distribution. With increasing redshift, this population progressively disappears from the CMASS sample.

At redshifts above 0.3, closer to the actual redshift range targeted by the CMASS selection cuts, the (very few) galaxies with detected emission lines are mostly AGN. Interestingly, the majority of these AGN are Seyfert. It should be emphasised, however, that AGN fractions derived through optical emission line ratios are critically dependent on S/N ratios. In particular LINERs, typically having lower emission line fluxes, tend to drop out of the sample as the S/N decreases. Hence, the relatively low fraction of BOSS galaxies with LINER emission at high redshifts is most likely a selection effect, and the effect of S/N ratio needs to be considered carefully in scientific analyses.

\begin{figure*}
\includegraphics[width=0.33\textwidth]{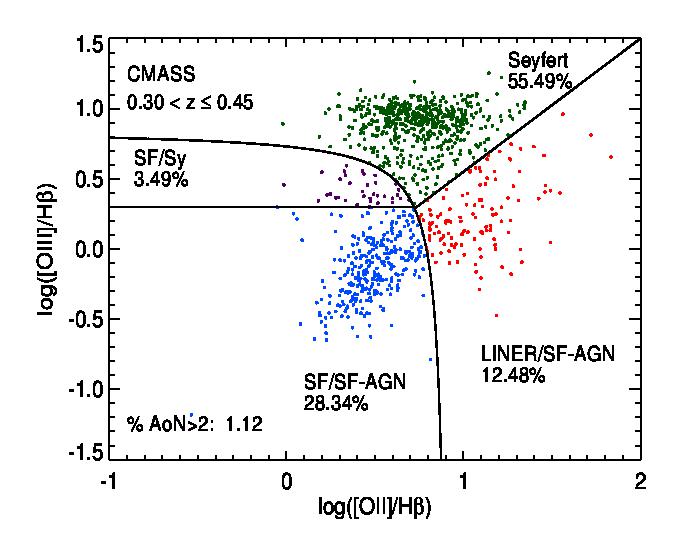}
\includegraphics[width=0.33\textwidth]{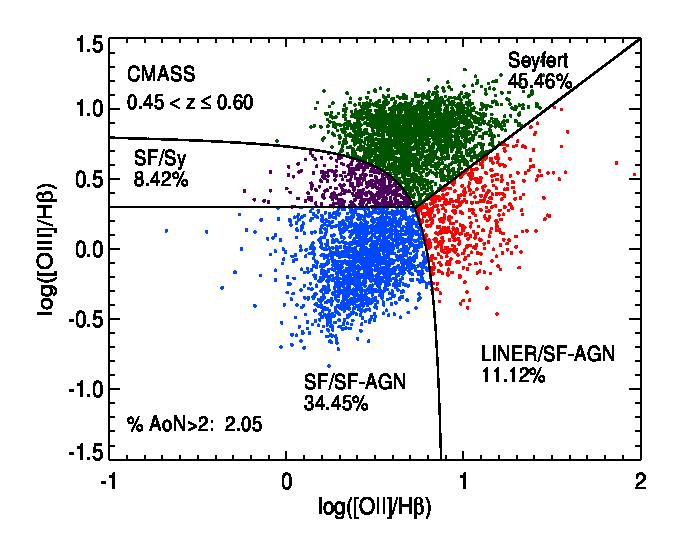}
\includegraphics[width=0.33\textwidth]{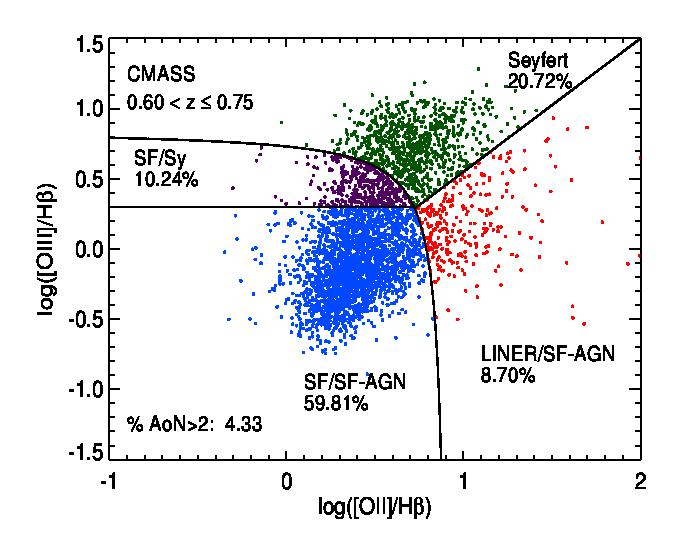}
\caption{Emission line classification \citep{Lamareille10} for CMASS BOSS galaxies at higher redshift bins (redshift increasing from left to right). The fraction of objects for which all three emission lines are detected (requiring the amplitude-over-noise ratio of all three lines to be larger than two) is given in the bottom-left corner of each panel. The fractions of galaxies in each of the emission line classes are given by the labels in the panels. The relative fractions of star forming and AGN in the redshift interval $0.3\le z\le 0.45$ agree well with the results form the BPT classification in Fig.~\ref{fig:bpt}. Beyond those redshifts, the fraction of star forming CMASS galaxies increases considerably with increasing redshift.}
\label{fig:lama}
\end{figure*}

\subsection{Emission line properties at higher redshifts}
\label{sec:lama}
{\bf\rm The analysis of star forming and AGN fractions at redshifts $z>0.45$ requires alternative methods, as the BPT diagnostic lines \Ha\ and [NII]$\lambda 6583$ are not within the observed wavelength range. Much of the separation between SF and (Seyfert-like) AGN is provided by the [OIII]/\Hb\ ratio, which is observable over the full BOSS redshift range. The main difficulty is the identification of LINER-like emission and SF-AGN composites with low [OIII]/\Hb\ where most star forming objects lie. These are characterised quite well by the by enhanced [NII]/\Ha\ ratios in the BPT diagram. Several alternative diagnostics have been discussed in the literature. \citet{Lamareille10} suggest to replace [NII]/\Ha\ by the blue emission line ratio [OII]/\Hb, which does help to identify LINER-like emission through enhanced [OII]/\Hb\ ratios. Alternative approaches use additional information from the stellar population properties of the host galaxy to separate AGN from star formation \citep{Juneau11,Yan11,Trouille11}. The Mass-Excitation method (MEx) by \citet{Juneau11} takes advantage of the fact that AGN are typically hosted by relatively massive galaxies, while \citet{Yan11} and \citet{Trouille11} could develop quite powerful ways to identify AGN through the relatively red $U-B$ and $g-z$ rest-frame colours of their host galaxies.

The MEx method only works over a large range in galaxy mass, hence cannot be applied to the BOSS sample which covers a very narrow range at the high-mass end \citep{Maraston12}. A potential problem of the approaches by \citet{Yan11} and \citet{Trouille11} is that we may not necessarily know how host galaxy properties change with redshift. By tapping into the host galaxy properties we may introduce a bias or some contamination in the classification calibrated at low redshifts. Also, the translation into rest-frame colours requires stellar population modelling, which introduces further uncertainties. In particular, the rest-frame z-band wavelength range is not covered by SDSS photometry at BOSS redshifts and requires near-IR imaging which we do not have available.

We therefore decided to focus on using the blue emission line diagnostics by \citet{Lamareille10}, which seems to be most comparable to the BPT approach used in Section~\ref{sec:bpt} for redshifts $z<0.45$. The major drawback is that, while the blue emission line diagnostic diagram does help to identify LINER-like emission through the enhanced [OII]/\Hb, the LINER region is still significantly contaminated with star forming objects and SF-AGN composites. This makes the identification of a pure LINER sample impossible. Also, SF-AGN composites cannot be identified. Still, the diagnostic is useful to identify overall fractions in star forming and (Seyfert-type) nuclear activity.

We analyse the emission line properties of CMASS galaxies in the redshift range $0.3<z<0.75$. In total there are 373,924 CMASS galaxies at $z>0.3$, for 10,238 of which (2.7 per cent) all three blue diagnostic emission lines ([OII]$\lambda\lambda 3726+2729$, \Hb, [OIII]$\lambda 5007$) could be detected at an AoN larger than 2. Fig.~\ref{fig:lama} presents the results for the redshift intervals $0.3< z<0.45$, $0.45<z<0.6$, and $0.6<z<0.75$, equivalent to the BPT diagnostic of Fig.~\ref{fig:bpt} at lower redshifts. The fractions of galaxies in each of the emission line classes are given by the labels in the panels.

The relative fractions of star forming/SF-AGN composite and AGN in the redshift interval $0.3\le z\le 0.45$ agree well with the results from the BPT classification in Fig.~\ref{fig:bpt}. Beyond those redshifts, the fraction of star forming CMASS galaxies increases considerably with increasing redshift. At redshifts above 0.6, almost two thirds of the CMASS galaxies with detected emission lines are star forming or SF-AGN composites. This trends is most probably due to the fact that the \Hb\ emission line is the weakest among those three, so that objects with low \Hb\ fluxes drop out first with increasing redshift. These are AGN, while star forming galaxies with stronger \Hb\ emission stay above the AoN threshold. A scientific analysis aimed at studying relative fractions of star forming and AGN with redshift will need to perform a careful assessment of this selection effect, which goes beyond the scope of this paper. Still, the BOSS sample is certainly useful for the identification and selection of massive star forming galaxies at redshifts around $z\sim 0.6$. {\bf\rm A first analysis of this kind has been carried out by \citet{Chen13} who study the link between radio AGN activity and star formation in CMASS galaxies around $z\sim 0.6$.}

}

\section{Target selection colour-colour space}
\label{sec:target}
\begin{figure*}
\centering\includegraphics[width=0.49\textwidth]{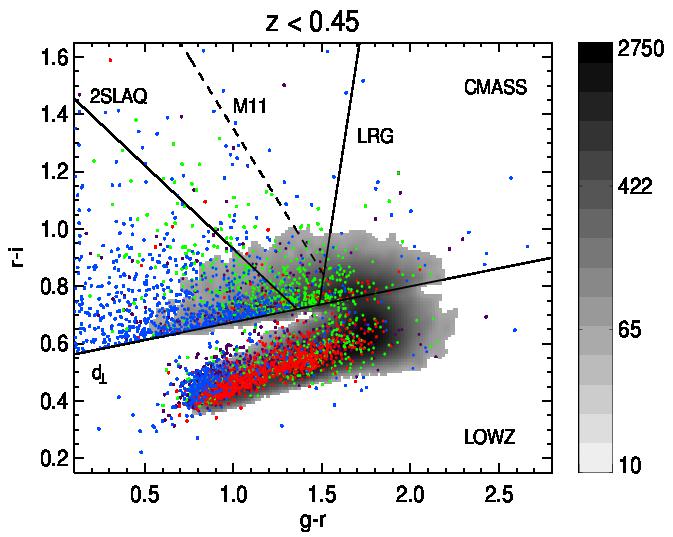}
\centering\includegraphics[width=0.49\textwidth]{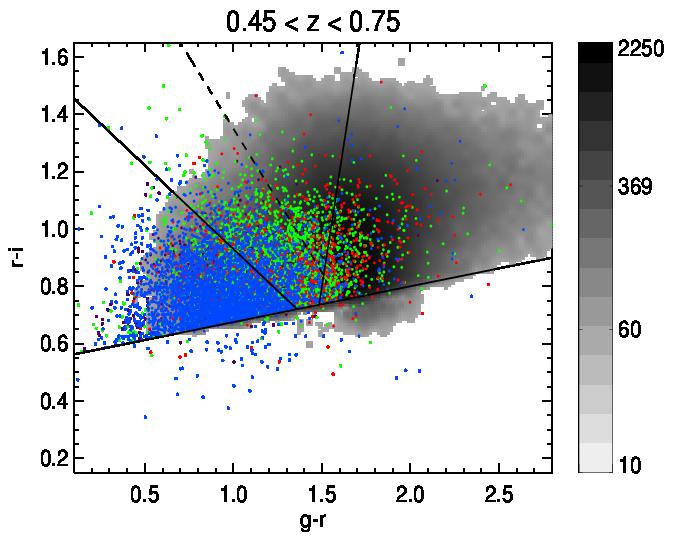}
\caption{\bf\rm BOSS target selection colour-colour diagram (based on galactic extinction-corrected {\em modelmags}). Objects below redshift $z=0.45$ and above $z=0.45$ are plotted in the left-hand and right-hand panels, respectively. Emission line classification is based on the BPT diagram of Fig.~\ref{fig:bpt} ($z<0.45$, left-hand panel) and the \citet{Lamareille10} diagnostics of Fig.~\ref{fig:lama} ($z>0.45$, right-hand diagram). Emission line free galaxies are plotted as grey density map. Galaxies with emission line detections at an amplitude-over-noise ratio larger than 2 in all diagnostic lines are overplotted as coloured symbols, following the colour-coding of Figs.~\ref{fig:bpt} and \ref{fig:lama}. The solid line labelled $d_{\perp}$ separates the high-$z$ sample CMASS from the low-$z$ sample LOWZ ($d_{\perp}\equiv (r-i)-(g-r)/8=0.55$). Galaxies above this line are generally at $z>0.4$. The solid lines labelled 'LRG' and '2SLAQ' are the selection cuts used in \citet{Eisenstein01} and \citet{Cannon06}. The dashed line is the dividing line between early-type and late-type from \citet{Masters11}. Total galaxy numbers and emission line class fractions are given in Table~\ref{tab:sections}.}
\label{fig:target}
\end{figure*}

We have shown that a small but significant fraction of BOSS galaxies contains emission lines, encompassing all ionisation classes from star forming to AGN. It is interesting to investigate, whether these emission line properties follow some distinct pattern in the colour-colour diagram used for target selection in BOSS. In the following we {\bf\rm split in the redshift ranges $z< 0.45$ and $0.45<z<0.75$, so that we can benefit from the full BPT classification where accessible for the separation between emission-line free galaxies, star forming and AGN. In Fig.~\ref{fig:target} we present the colour-colour plot $g-r$ vs $r-i$ of all DR9 BOSS galaxies. The left-hand panel is for $z<0.45$, and the emission line classes are derived form the BPT classification. The right-hand panel shows BOSS galaxies in the redshift range $0.45<z<0.75$, and the emission line classification is based on the blue diagnostic diagram of \citet{Lamareille10}.}

\subsection{\boldmath The ($g-r$) vs ($r-i$) diagram}

Galactic extinction-corrected {\em modelmags} are used. The solid line labelled $d_{\perp}$ indicates the major colour selection cut 
\begin{equation}
d_{\perp}\equiv (r-i)-(g-r)/8=0.55
\end{equation}
that has been used to separate the high-redshift CMASS and the low-redshift LOWZ samples \citep{Dawson12}. The colour $r-i$ is an excellent redshift indicator around redshifts of $z=0.4$, because the $4000$\AA\ break passes from the $g$-band into the $r$-band. As a consequence, galaxies become significantly redder in $r-i$ and slightly bluer in $g-r$ with increasing redshift beyond a redshift of $z=0.4$ \citep{Eisenstein01}. This pattern is displayed by the BOSS galaxy data in Fig.~\ref{fig:target}. Galaxies above the solid line ($d_{\perp}\ge 0.55$) define the CMASS sample and typically have redshifts $z\ga 0.4$, while everything below that line is in the LOWZ sample with typical redshifts of $z\la0.4$. The solid lines labelled 'LRG' and '2SLAQ' are the selection cuts used by \citet{Eisenstein01} to define the SDSS LRG sample and by \citet{Cannon06} to define the 2SLAQ LRG sample, respectively. The dashed line is the dividing line between early-type and late-type from \citet{Masters11}.

The correlation seen in the LOWZ sample at $z<0.45$ (left-hand panel) is a sequence of varying redshift in which both $g-r$ and $r-i$ become redder with increasing redshift up to $z=0.4$. The sequence is relatively tight, because redshift effects dominate the observed colours. The distribution of CMASS galaxies in the $g-r$ vs $r-i$ colour-colour plane is quite different. As the observed $g-r$ colour depends strongly on the star formation history of the galaxy, a wide range in $g-r$ colours is covered. The observed $r-i$ colour, instead, is generally red with a relatively small dynamical range and much less dependent on the star formation history. Passively evolving galaxies populate the top right-hand section of the diagram, as indicated by the LRG selection line.

\subsection{Correlations between observed frame colours and emission line properties}
The full BOSS sample is plotted as grey symbols, galaxies for which emission lines have been detected are plotted as coloured symbols. The various emission-line classes derived in the present work are shown using the colour-code of Fig.~\ref{fig:bpt} (left-hand panel) and Fig.~\ref{fig:lama} (right-hand panel). 

As is to be expected, there is a large overlap of the various emission line classes for the LOWZ sample at $z<0.45$. Still, galaxies with current star formation activity (blue points) are slightly shifted towards bluer $g-r$ colour by $\sim 0.1\;$mag, while no measurable offset is observed for the AGN populations (green and red symbols). Galaxies with SF-AGN composite spectra can be found in between those two extremes (purple points). This pattern of star-forming galaxies in the 'blue cloud' and transition objects in the 'green valley' between the blue cloud and the red sequence has previously been seen in colour-magnitude diagrams of SDSS-II galaxies \citep{Graves07,Schawinski07b,Salim07}.

For the CMASS sample at redshifts above $z\sim 0.4$, the separation of the various emission-line classes is far more pronounced. There are clear correlations between the observed frame colours and the emission-line properties of CMASS galaxies. There is a strong bias such that CMASS galaxies without emission lines are predominantly found in the classical LRG section towards the right-hand side of the diagram at red $g-r$ colours (most evident in the right-hand panel). CMASS galaxies with emission lines, instead, are more likely to be found at blue $g-r$ colour, and galaxies with star formation activity (blue points), in particular, populate the blue end of the $g-r$ colour space, typically having $(g-r)\la 1.0$. AGN, both Seyfert and LINER, preferentially occupy intermediate $g-r$ colour in between those two extremes. This pattern can be observed in both the low-$z$ and the high-$z$ versions of the diagram. It should be noted that a large fraction of the star forming galaxies in CMASS/BLUE at blue $g-r$ colours at $z<0.45$ (left-hand diagram) are low redshift ($z\sim 0.1$) interlopers that fell into the CMASS colour cut region most likely because of dust reddening as discussed in Section~\ref{sec:emission} (see Fig.~\ref{fig:bpt}). 

Table~\ref{tab:sections} quantifies the fractions of emission line galaxies and the various emission line classes as well the total number of objects for the four sections in Fig.~\ref{fig:target} defined by the solid lines:  CMASS/LRG (above $d_{\perp}$, right of 'LRG'), CMASS/M11 (above $d_{\perp}$, between 'LRG' and 'M11'), CMASS/2SLAQ (above $d_{\perp}$, between 'M11' and '2SLAQ'), CMASS/BLUE (above $d_{\perp}$, left of '2SLAQ'), LOWZ (below $d_{\perp}$). {\bf\rm CMASS/BLUE is most abundant in emission-line objects (23 per cent at $z<0.45$ and 7 per cent at $z>0.45$) and star forming galaxies (around half of emission line spectra). The classical LRG section (CMASS/LRG), instead, is devoid of emission line objects (less than per cent). CMASS/2SLAQ, instead, contains the largest fraction of galaxies with Seyfert-like AGN emission lines (60 per cent).}

\begin{table}
\caption{Percentages of emission line classes from Fig.~\ref{fig:target}.}
\begin{tabular}{rrrrrr}
\hline\hline
 & CMASS & CMASS & CMASS & CMASS & LOWZ\\
 & LRG & M11 & 2SLAQ & BLUE & \\
\hline
\multicolumn{6}{l}{$z<0.45$ (left-hand panel in Fig.~\ref{fig:target})}\\
\hline
Total & 17,390 & 518 & 8,685 & 6,890 & 104,317\\
Emission & 0.84 & 6.76 & 4.17 & 23.35 & 2.46\\
SF & 19.86 & 42.86 & 17.96 & 69.48 & 12.94\\
SF/AGN & 13.70 & 14.29 & 15.19 & 10.69 & 36.18\\
Seyfert & 53.42 & 40.00 & 61.60 & 17.65 & 15.59\\
LINER & 13.01 & 2.86 & 5.25 & 2.18 & 35.28\\
\hline
\multicolumn{6}{l}{$0.45<z<0.75$ (right-hand panel in Fig.~\ref{fig:target})}\\
\hline
Total & 211,389 & 55,220 & 53,609 & 28,777 & 4,085\\
Emission & 0.33 & 0.79 & 4.78 & 18.95 & 3.30\\

SF & 15.20 & 12.79 & 26.86 & 60.22 & 60.74\\
SF/AGN & 3.55 & 2.74 & 6.95 & 11.57 & 5.19\\
Seyfert & 50.43 & 65.98 & 53.69 & 22.38 & 17.04\\
LINER & 30.82 & 18.49 & 12.50 & 5.83 & 17.04\\
\hline
\end{tabular}
\label{tab:sections}
\end{table}

\subsection{Star formation histories and morphologies}
The dependence of star formation activity on observed $g-r$ colour for the CMASS galaxies as found in the present work is not surprising. As discussed above, it was to be expected that the blue $g-r$ colour is driven by the presence of young stellar populations caused by recent or current star formation episodes. \citet{Tojeiro12} analyse the star formation histories of CMASS galaxies through spectral SED fitting and find extended star formation in blue CMASS galaxies \citep[see also][]{Chen12}. The present results confirms this finding.

If the CMASS galaxies with blue $g-r$ are dominated by young stellar populations and have a fraction of star forming galaxies, one would expect to find mostly late-type galaxies. This is indeed the case. \citet{Masters11} analyse the morphologies of BOSS galaxies through HST/COSMOS imaging and find that most CMASS galaxies with blue $g-r$ colour are in fact late type systems. Their morphology-driven dividing line between early-type and late-type at $g-i=2.55$ (dashed line in Fig.~\ref{fig:target}) separates quite well between star forming and passive galaxies.

\section{Conclusions}
\label{sec:conclusions}
BOSS is one of four surveys of the SDSS-III collaboration using an upgrade of the multi-object spectrograph on the 2.5m SDSS telescope to collect spectra of galaxies and quasars over 10,000 deg2 on the sky \citep{Eisenstein11}. BOSS has started operation in autumn 2009 and by 2014 it will have observed about 1.5 million luminous galaxies up to redshifts $z\sim 0.7$. Targets are selected from SDSS imaging split in a high redshift sample called CMASS and a low redshift sample LOWZ. BOSS multi-object spectroscopy is being performed with an upgrade of the SDSS fibre-fed spectrograph providing spectra of reasonable signal-to-noise ratio in 1-hour exposures down to the limiting magnitude in the i-band of 19.9 mag. 

We perform a spectroscopic analysis of 492,450 galaxy spectra that are part of the ninth SDSS data release in July 2012, the first public data release of BOSS spectra \citep{SDSSDR9}. We use the publicly available codes pPXF \citep{CE04} and GANDALF v1.5 \citep{Sarzi06} to calculate stellar velocity dispersions and to derive emission line fluxes and equivalent widths. The new stellar population models from \citet{Mastro11} based on the MILES stellar library \citep{Sanchez06a} are adopted. To calibrate the procedure, we have used our technique to derive stellar velocity dispersions and emission line properties for a subset of SDSS galaxies from Data Release 7 \citep{SDSSDR7} and found satisfying agreement. The velocity dispersions are in good agreement with a small median offset in $\sigma$ of 2~km~s$^{-1}$ at a dispersion of 30~km~s$^{-1}$. Also the emission line measurements generally agree reasonably well showing tight correlations with small scatter and only small offsets. Our measurements of emission line fluxes and EWs tend to be slightly larger than in DR7 by $\sim 0.1\;$dex with a dispersion of $\sim 0.2\;$dex. Still, the comparison is satisfying overall, and residual discrepancies will most likely be caused by differences in the treatment of reddening, absorption line correction, and continuum fitting.

We show that the typical signal-to-noise ratio of BOSS spectra, despite being low, is sufficient to measure simple dynamical quantities such as stellar velocity dispersion for individual objects. We verify the reliability of our measurements on individual BOSS spectra through comparison with high signal-to-noise spectra from repeat-plate observations in BOSS using a sub-sample of 574 BOSS galaxies mainly from plates 3615 and 3647 with at least six 1-hour observations each. The agreement is very good. There is no systematic offset between measurements on individual (low S/N) and stacked (high S/N) spectra. We also do not find any systematic trend with S/N ratio. Finally, we compare our measurements with independent measurements within the BOSS collaboration by \citet{Bolton12} and \citet{Chen12}, and find good agreement. The typical error in the velocity dispersion measurement is 14 per cent, and 93 per cent of BOSS galaxies have velocity dispersions with an accuracy better than 30 per cent. We show that the typical velocity dispersion of a BOSS galaxy is $\sim 240$ km s$^{-1}$. The distribution in velocity dispersion is nearly Gaussian and is redshift independent between redshifts 0.15 and 0.7. At redshifts below $z\sim 0.15$ and above $z\sim 0.75$, the distributions are skewed and slightly offset toward lower and higher velocity dispersions, respectively. This reflects the survey design targeting massive galaxies with an approximately uniform mass distribution in the redshift interval $0.15<z<0.75$.

We show that emission lines can be measured on BOSS spectra, but the majority of BOSS galaxies lack detectable emission lines, as is to be expected because of the target selection design toward massive galaxies.

We analyse the emission line properties for a subsample of 140,596 galaxies below $z=0.45$, so that the full set of key diagnostic emission lines  H$\beta$, [OIII]$\lambda 5007$, H$\alpha$, and [NII]$\lambda 6583$ {\bf\rm is accessible in the rest-frame spectra}. {\em All four} diagnostic lines are detected at an amplitude-over-noise ratio above two for 4,887 spectra (3.5 per cent). For these, we present the classical diagnostic diagrams \citep{BPT81} to divide star-forming objects from AGN separately for the high-$z$ sample CMASS and the low-$z$ sample LOWZ. We find that the emission line properties are strongly redshift dependent. Furthermore, there is a clear correlation between observed frame colours and emission line properties. In general, the fraction of star forming galaxies decreases and the fraction of AGN increases with increasing redshift, mostly owing to selection effects. Within in the LOWZ sample, the majority of emission-line galaxies has some AGN component, the fraction of purely star forming galaxies only being a few per cent at $z>0.15$. The CMASS sample, instead, contains bluer galaxies and the fraction of star forming galaxies is as high as 20 per cent at $z>0.3$. Interestingly, there are some CMASS galaxies at low redshifts ($z<0.15$) that are star forming and fell into the CMASS colour selection cut most probably due to dust reddening.

{\bf\rm To assess the emission line properties of BOSS galaxies at higher redshifts, we additionally study the 373,924 CMASS galaxies in the redshift range $0.3<z<0.75$, containing 10,238 objects  (2.7 per cent) with significant emission line detections. For this purpose we use the blue diagnostic diagram of \citet{Lamareille10} based on the emission lines [OII]$\lambda\lambda 3726+3729$, H$\beta$, and [OIII]$\lambda 5007$.  For this sample, the fraction of star forming galaxies is considerably higher, which is most probably due to the fact that the \Hb\ emission line is the weakest among the three diagnostic lines, so that objects with low \Hb\ fluxes, hence AGN, drop out first with increasing redshift. Therefore, the BOSS sample turns out to be instrumental for the identification and selection of massive star forming galaxies at redshifts around $z\sim 0.6$.}

Finally, we show that CMASS galaxies whose emission lines are produced by star formation activity have blue observed $g-r$ colours and are well separated in the $g-r$ vs $r-i$ target selection diagram. 

To conclude, BOSS offers spectra of a large sample of galaxies up to redshifts $\sim 0.8$. The quality of BOSS spectroscopy, even though designed for redshift determination, allows the measurement of simple quantities on individual BOSS spectra for a wealth of galaxy evolution studies on dynamical, gas, and stellar population properties.

\section*{Acknowledgements}
The Science, Technology and Facilities Council is acknowledged for support through the 'Survey Cosmology and Astrophysics' rolling grant, ST/I001204/1. Numerical computations were done on the Sciama High Performance Compute (HPC) cluster which is supported by 
the ICG, SEPnet and the University of Portsmouth.

Funding for SDSS-III has been provided by the Alfred P. Sloan Foundation, the Participating Institutions, the National Science Foundation, and the U.S. Department of Energy Office of Science. The SDSS-III web site is http://www.sdss3.org/.

SDSS-III is managed by the Astrophysical Research Consortium for the Participating Institutions of the SDSS-III Collaboration including the University of Arizona, the Brazilian Participation Group, Brookhaven National Laboratory, University of Cambridge, Carnegie Mellon University, University of Florida, the French Participation Group, the German Participation Group, Harvard University, the Instituto de Astrofisica de Canarias, the Michigan State/Notre Dame/JINA Participation Group, Johns Hopkins University, Lawrence Berkeley National Laboratory, Max Planck Institute for Astrophysics, Max Planck Institute for Extraterrestrial Physics, New Mexico State University, New York University, Ohio State University, Pennsylvania State University, University of Portsmouth, Princeton University, the Spanish Participation Group, University of Tokyo, University of Utah, Vanderbilt University, University of Virginia, University of Washington, and Yale University.



\bsp
\label{lastpage}

\end{document}